\documentclass[a4paper,11pt]{article}
\pdfoutput=1
\usepackage{jheppub}
\usepackage{amsmath,amssymb,amsfonts,bm,amscd,slashed}
\usepackage{graphicx}
\usepackage{mathtools}
\usepackage{mathrsfs}

\newcommand{\bea}{\begin{eqnarray}}
\newcommand{\eea}{\end{eqnarray}}
\newcommand{\be}{\begin{equation}}
\newcommand{\ee}{\end{equation}}

\newcommand{\Z}{{\mathbb Z}}
\newcommand{\R}{{\mathbb R}}
\newcommand{\C}{{\mathbb C}}

\newcommand{\cN}{{\mathcal{N}}}

\newcommand{\pD}{{\mathscr D}}
\newcommand{\bB}{{\mathscr B}}
\newcommand{\dd}{{\rm d}}

\def\tilde{\widetilde}
\def\hat{\widehat}
\def\bar{\overline}

\def\cD{{\mathcal D}}

\def\cG{{\mathcal G}}
\def\cH{{\mathcal H}}

\def\cL{{\mathcal L}}

\def\cN{{\mathcal N}}

\def\cP{{\mathcal P}}

\def\qft{{\text{QFT}}}

\renewcommand{\bar}{\overline}
\renewcommand{\hat}{\widehat}

\preprint{CALT-TH 2018-024}

\title{Gluing I: Integrals and Symmetries}

\author[1]{Mykola Dedushenko}

\affiliation[1]{Walter Burke Institute for Theoretical Physics,\\ California Institute of Technology,\\ Pasadena, CA 91125, USA}

\emailAdd{dedushenko@gmail.com}

\abstract{We review some aspects of the cutting and gluing law in local quantum field theory. In particular, we emphasize the description of gluing by a path integral over a space of polarized boundary conditions, which are given by leaves of some Lagrangian foliation in the phase space. We think of this path integral as a non-local $(d-1)$-dimensional \emph{gluing theory} associated to the parent local $d$-dimensional theory. We describe various properties of this procedure and spell out conditions under which symmetries of the parent theory lead to symmetries of the gluing theory. The purpose of this paper is to set up a playground for the companion paper where these techniques are applied to obtain new results in supersymmetric theories.}

\begin{document}
	\maketitle
	\flushbottom



\section{Introduction}
\begin{figure}[t!]
	\centering
	\includegraphics[width=1\textwidth]{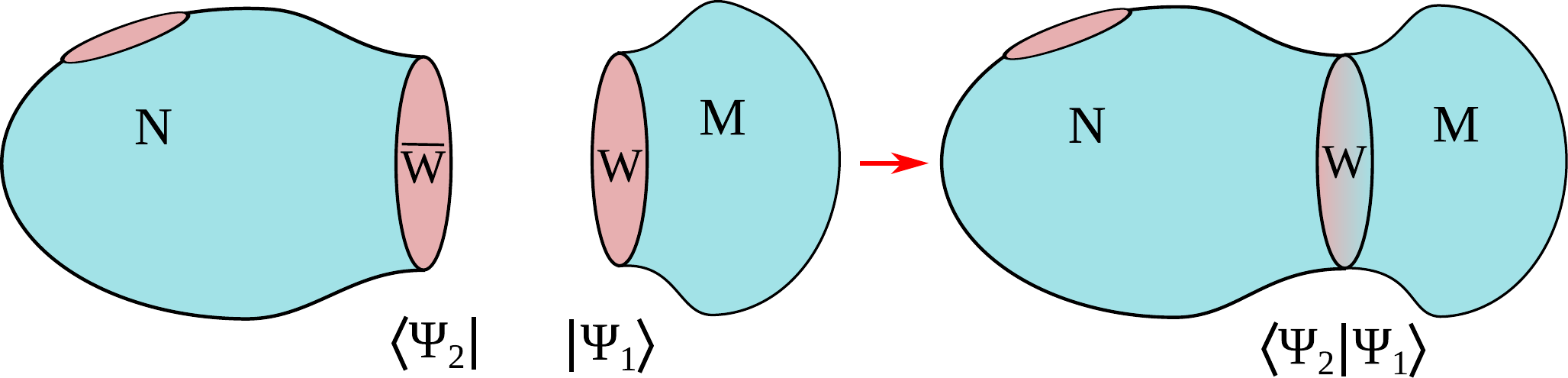}
	\caption{\label{fig:glue} An illustration of gluing and associated composition of boundary states.}
\end{figure}

Local quantum field theories are our main theoretical tool in high energy physics. Their distinguishing property---locality---can be formulated elegantly, at least in Euclidean signature, as a cutting and gluing axiom stating that one can consistently cut and glue manifolds with QFTs living on them. More precisely, if we have two $d$-manifolds $M$ and $N$ with common boundary component $W$:
\begin{equation}
W\subset \partial M,\quad \bar{W} \subset \partial N,
\end{equation}
where $\bar{W}$ denotes $W$ with opposite orientation, they can be attached along $W$, and this topological operation has a counterpart for the QFT data.\footnote{We will usually assume the manifolds to be smooth, both before and after gluing, see more comments on this later.} In topology, the gluing operation (or the fibered coproduct) is denoted by:
\begin{equation}
N\cup_W M = \left(M \coprod N\right) \Big{/}\sim\, ,
\end{equation}
where $\sim$ is the equivalence relation identifying points of $W$ with those of $\bar{W}$. A $d$-dimensional QFT, denoted as $\qft_d$, associates to $W$ a space of states $\cH_W$ (a possibly infinite dimensional topological vector space) and its dual $\cH_W^\vee$ to $\bar{W}$. The dynamics of quantum fields living in the bulk of $M$ and $N$ produces boundary states, which are described by (co)vectors from the corresponding spaces. We denote them using the standard bra- and -ket symbols:
\begin{equation}
\label{BStates}
|\Psi_1\rangle \in \cH_W,\quad \langle \Psi_2| \in \cH_W^\vee.
\end{equation}
All physical information about the bulk dynamics of $M$ and $N$ that is accessible at the boundary $W$ is encoded in these states. On the other hand, for a given QFT, the space $\cH_W$ depends only on the geometry of the infinitesimal neighborhood of $W$ and not on any details of the deep bulk,\footnote{A more precise way to say it is that $\cH_W$ depends on the germ $[W]$ of $W\subset N\cup_W M$.} which are encoded in $|\Psi\rangle\in \cH_W$. This is one key assumption in local QFTs, which further leads to the main axiom, the gluing or sewing property, stating that the gluing operation corresponds to the composition of boundary states:
\begin{equation}
N \cup_W M \longleftrightarrow \langle\Psi_2|\Psi_1\rangle,
\end{equation}
as also illustrated in Figure \ref{fig:glue}. The opposite operation is cutting: if a QFT is defined on $N\cup_W M$, we can cut it along $W$, and the bulk dynamics will generate unique boundary states as in \eqref{BStates}. The locality postulates that these two operations are exactly opposite to each other: one can construct QFT on a bigger space by gluing it from smaller pieces. 

The cutting and gluing axiom is closely related to linearity in quantum mechanics (in which case locality postulates existence of the Hilbert space and implies existence of the Hamiltonian) and goes back to Dirac and the resolution of the identity trick, $1=\int |q\rangle\,\dd q\,\langle q|$ \cite{Dirac:1927we,dirac1967principles}. This later led to the discovery of Feynman's path integrals \cite{Feynman:1948ur}, which for local theories can be deduced by cutting spacetime into infinitely many small pieces and then gluing them back together. Skipping several decades ahead, the idea of cutting and gluing has become the base of Segal's approach to conformal field theories, which then took several more decades to be published \cite{Segal:1987sk,Segal:1988zk,segal_2004}. Soon after this philosophy has emerged, and after the discovery of topological field theories \cite{Schwarz:1978cn,Witten:1988ze}, the cutting and gluing law (or ``the sewing law'') played a central role in the axiom system defining the mathematical notion of TQFT \cite{Atiyah1988}. The main observation was that if one abstractly considers assignment of spaces of states to $(d-1)$-manifolds $W$, the cutting and gluing property can be seen as a statement that this assignment is a functor from the category of appropriate cobordisms to the category of vector spaces. In the topological case, this was later expanded to include higher categories of cobordisms, which lead to the notion of extended TQFTs \cite{Baez:1995xq} (see also \cite{Freed:2008jq,Freed:2009qp}) and their classification \cite{2009arXiv0905.0465L}. Recently, there has been a number of works on general perturbative quantization of (not necessarily topological) gauge theories on manifolds with boundary, where the so-called BV-BFV formalism compatible with cutting and gluing was developed \cite{Cattaneo:2012qu,Cattaneo:2012zs,Alekseev:2012wc,Cattaneo:2015vsa,Cattaneo:2016hqk,Cattaneo:2017tef,Iraso:2018mwa}. More generally, this is part of a bigger program of functorial quantum field theory \cite{segal_roles}, which is a leading candidate (along with parallel progress in algebraic QFT) for the mathematical definition of QFT (see also \cite{stolz_notes} for some related discussions).

A lot of developments in quantum field theory rely in one way or another on the gluing property, many of which are classics, such as the operator-state correspondence in CFT, or the crossing symmetry (which can be thought of as equating cutting-and-gluing in $s$ and $t$ channels) \cite{Belavin:1984vu}, etc. Any attempt to fully review this subject would require including thousands of references, which by itself requires a hardly justifiable amount of work, at least within the scope of this paper. Therefore, we choose to focus on a particular aspect of gluing that will be detailed soon. 

Note that the standard TQFT axioms imply that the space of states is finite-dimensional. In this case, $\langle \Psi_1| \Psi_2\rangle$ is understood as a straightforward pairing between finite-dimensional vector spaces. In addition, given that manifolds can be glued topologically, gluing in TQFT is also well-defined. Both of these properties become more subtle once we depart from the world of topological theories (or, more precisely, from the standard TQFT axiomatics). Already in quantum mechanics, the space of states can be infinite-dimensional: typically, it is a functional space, in which case the inner product is still well-defined, $\langle\Psi_1|\Psi_2\rangle=\int\langle\Psi_1|q\rangle\,\dd q\,\langle q|\Psi_2\rangle$. Moving to QFTs in dimension $d>1$, the ``infinite-dimensionality'' of the space of states grows, and it becomes more difficult to make sense of $\langle\Psi_1|\Psi_2\rangle$. Explaining some properties of the latter is the main focus of the current paper. For completeness, we should say a few more words about other subtleties of gluing in non-TQFTs.

In non-topological theories, manifolds usually become equipped with smooth metric (and, possibly, other geometric structures), either Riemannian or pseudo-Riemannian (i.e., of Euclidean or Lorentzian signature). Therefore, not any pair of manifolds that can be glued topologically are allowed to be glued geometrically. It might happen that even though $M$ and $N$ have a common boundary component $W$, the metrics on $M$ and $N$ are incompatible: simply equipping $N\cup_W M$ with a ``sewed'' metric results in some sort of discontinuity at $W$ that is forbidden in the class of geometries that we are considering. In such cases, one could imagine smoothing out this discontinuity in the infinitesimal neighborhood of $W$, and it would be interesting to understand whether this can always be done and if it leaves an imprint on the resulting theory (in the form of some defect). This problem will not be relevant for us in this paper, and we will simply consider cases where the metrics on $M$ and $N$ are compatible, so that the ``sewed'' geometry is smooth. This is general enough because, as we mentioned, the space $\cH_W$ really depends on the germ of $W$ inside the $d$-dimensional space \cite{segal_roles}, not on $W$ alone.

Another problem that becomes apparent in a non-topological setting is that it is necessary to consider both Riemannian and pseudo-Riemannian geometries. From the physics point of view, the spaces $\cH_W$ and $\cH_W^\vee$ associated to $W$ and $\bar{W}$, as well as the ``gluing'' $\langle \Psi_2|\Psi_1\rangle$, are more naturally described in terms of a QFT on a pseudo-Riemannian space $W\times \R_t$. It is equipped with the metric $\dd s^2_W - \dd t^2$, where $t$ is a ``time'' coordinate on $\R_t$, and $\dd s^2_W$ is a Riemannian metric on $W$. This is usually referred to as ``quantizing on $W$''. On the other hand, one way to generate physical states in $\cH_W$ is by putting a theory on some arbitrary manifold $M$ with $\partial M=W$. The path integral on $M$ determines a boundary state in $\cH_W$ as a functional of boundary conditions. Close to the boundary, we could approximate $M$ by $W\times \R_t$, but globally $M$ might not possess a pseudo-Riemannian metric. Therefore, in general $M$ can only be a Riemannian manifold. This raises a practical concern: how do we describe a state living at the boundary of a Riemannian manifold as an element of the space $\cH_W$, given that $\cH_W$ is defined using the pseudo-Riemannian geometry $W\times \R_t$? It turns out that while in the simplest cases the answer is trivial---the boundary state, described as a functional, is not sensitive to the signature of metric on $\R_t$---in a more general situation, the answer involves certain analytic continuation. This subtlety will be of some importance to us, thus more details will be provided in the following sections.

The standard strategy to describe a boundary state at $W$ is by imposing various boundary conditions on quantum fields near $W$ and studying how the answer depends on them. For example, in the situation of Figure \ref{fig:glue}, the QFT on $M$ with fixed boundary conditions $\bB$ at $W$ associates a number to $M$, the partition function. For Lagrangian field theories, we assume, at least formally, that this number is given by a path integral over all field configurations on $M$ satisfying the boundary conditions $\bB$. This number is some functional of $\bB$, and we characterize the boundary state by this functional:
\begin{equation}
|\Psi_1\rangle \leftrightarrow \Psi_1[\bB].
\end{equation}
To fully determine this functional, that is to know how to evaluate it for arbitrary boundary conditions, it is enough to know $\Psi_1[\bB]$ for a certain complete family of boundary conditions, which is big enough but does not need to include all possible boundary conditions. A fairly general way to determine such families is by choosing a polarization $\cP$ on the phase space $X[W]$ of our QFT quantized on $W$. The latter is the formal phase space of the quantum-mechanical sigma model whose target space is the infinite-dimensional fields space ${\rm Fields}[W]$. The complete family of boundary conditions is given by the family of integral leaves of polarization $\cP$. The Lagrangian leaves from this family determine consistent boundary conditions, and we call them \emph{polarized} boundary conditions. So the boundary states, in general, can be formally characterized by functionals of polarized boundary conditions. There can be many choices of polarization, hence many different ways to represent boundary states as functionals. Denoting the boundary conditions imposed at $W$ and $\bar{W}$ by $\langle\bB|$ and $|\bB\rangle$ respectively, we can formally write:
\begin{equation}
\Psi_1[\bB]=\langle\bB|\Psi_1\rangle,\quad \Psi_2[\bB]=\langle\Psi_2|\bB\rangle.
\end{equation}
The gluing procedure can be represented as integration over the space of polarized boundary conditions (space of Lagrangian leaves) for a given polarization:
\begin{equation}
\label{gluingPI}
\langle\Psi_2|\Psi_1\rangle = \int_{\text{polarized }\bB} \pD \bB\, \Psi_2[\bB] \Psi_1[\bB].
\end{equation}
In spacetime dimension $d>1$, this integration is infinite-dimensional, thus is formal and requires regularization, just like the usual bulk path integral (in fact, the gluing integral inherits regularization from the bulk, see the next subsection).

Polarized boundary conditions can be described in terms of a set of independent fields on $W$ (for example, these could be restrictions of certain fields on $M$ to the boundary). This makes it very suggestive to interpret the path integral in \eqref{gluingPI} as a quantum field theory on $W$. In fact, this might be the unique way to interpret it: indeed, the only context in which we encounter path integrals is quantum field theory, and it might be the case that they simply do not exist outside of its framework, in a sense being synonymous to QFT. Within the scope of this paper, we simply think of \eqref{gluingPI} as a $(d-1)$-dimensional QFT on $W$, and we call it \emph{the gluing theory}. Therefore, we have the following correspondence:\footnote{One could try to generalize it further to a cascade of theories of the form $\qft_d \to \qft_{d-1} \to \qft_{d-2} \to \dots$. Such generalization is expected if we perform further cuts, resulting in manifolds with corners of various codimension, while still using local boundary conditions. In this paper, we are not trying to make any sense of it and only study the first gluing theory $\qft_{d-1}$.}
\begin{equation}
\qft_d \longrightarrow \text{the gluing theory } \qft_{d-1}.
\end{equation}

It is clear, however, that in general, this gluing $\qft_{d-1}$ cannot exist by itself as a $(d-1)$-dimensional theory, -- it can only be defined as a device sewing two $d$-dimensional theories. One obvious reason is that the integrand in \eqref{gluingPI} depends on states $\Psi_1$ and $\Psi_2$. Therefore, it would not make sense to consider $\qft_{d-1}$ on a separate $(d-1)$-manifold $W$, it always should be an interface connecting two $d$-manifolds, and the $d$-dimensional physics determines the integrand $\Psi_1[\bB]\Psi_2[\bB]$ of the $(d-1)$-dimensional path integral. Moreover, as we will see later, the gluing theory might appear anomalous: it might happen that the measure $\pD\bB$ is not invariant under global or even gauge symmetries. As we will see very explicitly in the next section, the gluing theory still makes perfect sense even in such situations due to the version of the anomaly inflow: the non-invariance of ``$\pD \bB$'' is canceled by the non-invariance of $\Psi_2[\bB]\Psi_1[\bB]$ whenever we know that the $d$-dimensional theory is non-anomalous.

\subsection{Addressing possible concerns}\label{sec:concerns}
Is it possible to compute the path integral in \eqref{gluingPI}, or even make any sense of it? At this level of generality, the problem seems hopeless: just to write the integrand of \eqref{gluingPI} already requires solving $\qft_d$ on $M$ with arbitrary polarized boundary conditions $\bB$. Furthermore, gluing by the boundary path integral, as described above, is a purely formal procedure which ignores that the boundary conditions are quantum objects: they can receive quantum corrections, get renormalized, and undergo non-trivial RG flows. The latter set of problems is very familiar already in the case of bulk path integrals: it can be concisely stated as the need for regularization in QFT. Therefore, in order to make any precise sense of \eqref{gluingPI}, one should assume the existence of good regularization. The most convenient choice is to ``induce'' regularization from the bulk path integral. Indeed, to define the bulk QFT, one already has to assume certain regularization. It would be quite unnatural to first remove that regularization and then reintroduce another regularization for gluing. Instead, one should perform gluing in the regularized theory. Vaguely speaking, this requires that the bulk regularization preserves locality, so that the gluing property holds in the regularized theory. The most canonical examples of such locality-preserving regularizations are lattice regularizations: once we assume that the bulk theory has been properly discretized, with the appropriate counterterms included (to ensure the existence of the continuum limit), the gluing integral becomes naturally discretized as well. It should be possible to adapt other physical regularizations to gluing as well. We will work under the assumption that the locality-preserving regularization exists.

With such an assumption, we can study certain general properties of \eqref{gluingPI} semiclassically, and under favorable conditions, even evaluate it exactly. One thing we can definitely tell about the gluing theory based on the semiclassical analysis is its field content. After fixing the family of polarized boundary conditions $\bB$ over which we wish to integrate, we know what fields on $W$ constitute this $\bB$. We will focus on theories whose UV Lagrangians are renormalizable and have no more than two derivatives, so the boundary fields will usually be given by restrictions of $d$-dimensional fields to the boundary, their normal derivatives, or various mixtures thereof.

One case when \eqref{gluingPI} can be evaluated exactly is for free theories, where it is possible to explicitly write the non-local action of $\qft_{d-1}$. This was first studied for bosons on Riemann surfaces in \cite{Morozov:1988xk, Losev:1988ea, Morozov:1988gj, Morozov:1988bu}. Recently, similar analysis was also performed for a free Maxwell theory (in arbitrary spacetime dimension) and 2d Yang-Mills in \cite{Blommaert:2018oue}. Another case in which one can efficiently evaluate the $\qft_{d-1}$ path integral is in the presence of extra symmetries, such as supersymmetry: under favorable conditions, it allows to make sense of \eqref{gluingPI} using the supersymmetric localization.

\subsection{What we really do}
This paper is designed to be a preparation for the companion paper \cite{glue2} and mostly reviews known facts, albeit from a somewhat unfamiliar perspective. After reviewing necessary facts on symplectic geometry of quantum mechanics, we emphasize the point of view which is useful for applications in \cite{glue2}: that gluing can be represented by a (generally non-local) $\qft_{d-1}$ on the gluing interface whose fields parametrize polarized boundary conditions of $\qft_d$. Though such a point of view is not strikingly new, its systematic study, to the best of author's knowledge, has not appeared in the literature. Also, very importantly, this approach holds more generally than special cases of gluing that appeared previously throughout the literature, such as, e.g., ``gluing by gauging''.\footnote{This refers to the following: gluing along $W$ with the Dirichlet boundary conditions on the two sides is done by gauging ${\rm diag}(G\times G)$, where $G$ is the global part of gauge transformations preserved by the Dirichlet boundary condition. Such prescription only works in pure gauge theories, because in this case $\qft_{d-1}$ is a theory of a single gauge field. For more general gauge theories, $\qft_{d-1}$ would also include various matter fields. Gluing in the presence of matter is often done in the literature by adding boundary potentials.}

We address the question of symmetries in $\qft_{d-1}$, namely, when non-anomalous symmetries of $\qft_d$ induce symmetries of $\qft_{d-1}$. The answer turns out to be quite simple and expected, though important for the applications in \cite{glue2}, and is called \emph{The Main Lemma} in Section \ref{sec:gluesym}. As we will see, the symmetry of $\qft_d$ descends to $\qft_{d-1}$ if the two conditions are satisfied: first, this symmetry should preserve the polarization that was used in the definition of $\qft_{d-1}$; second, the states $\Psi_1$ and $\Psi_2$ that we want to glue should be annihilated by the corresponding charge. We will prove this for quantum mechanics in the next section. So far, we can simply say that the symmetry should preserve the family of polarized boundary conditions, namely transform one boundary condition into another inside the same family. Under this assumption, the gluing theory acquires a symmetry.

Interesting simplifications might occur in supersymmetric theories. Applying the Main Lemma to supersymmetry shows that whenever the polarization is preserved by SUSY $Q$ of the parent theory $\qft_d$, the gluing theory $\qft_{d-1}$ describing the convolution of $Q$-closed states $\Psi_{1,2}$ becomes supersymmetric. In certain cases, we can further apply supersymmetric localization and reduce the path integral in \eqref{gluingPI} to a finite dimensional integral over the space of supersymmetric boundary conditions. This allows to derive a number of interesting formulas describing gluing of supersymmetric theories on manifolds with boundary. We call them ``the gluing formulas''. As was recently shown in \cite{Dedushenko:2017avn}, such formulas can be quite useful for various problems in SUSY field theories. For example, a certain gluing formula for 3D $\cN=4$ theories on spheres and hemispheres was used in \cite{Dedushenko:2017avn} as a tool to describe their quantized Coulomb branches. Such supersymmetric applications of the formalism developed in the current paper are discussed in a companion paper \cite{glue2}. Speaking more broadly, the general idea of gluing has appeared in multiple references on supersymmetric field theories, see e.g. \cite{Drukker:2010jp, Gaiotto:2014gha, Dimofte:2011py, Beem:2012mb, Gang:2012ff, Gadde:2013wq, Pasquetti:2016dyl, Bullimore:2014nla, Hori:2013ika, Gadde:2013sca, Honda:2013uca, Cabo-Bizet:2016ars, Gava:2016oep, Gukov:2017kmk, LeFloch:2017lbt, Bawane:2017gjf, Dimofte:2017tpi}.

\subsection{The structure of this paper}

The structure of this paper is as follows. In Section 2 we start by describing the gluing formalism for the simplest case of Quantum Mechanics. We review some basic notions of geometric quantization, such as polarization, and show explicitly how this leads to the gluing theory (which is a zero-dimensional QFT in this case). We then prove that every symmetry of the parent theory (quantum mechanics in this case) that preserves polarization and the two boundary states descends to the symmetry of the gluing theory. Thinking of $\qft_d$ on $W_{d-1}\times \R$ as a quantum mechanics $\R\to \text{Fields}[W_{d-1}]$, we extend the statements in Section 2 to the case of higher-dimensional quantum field theories (modulo caveats related to regularization as mentioned in Section \ref{sec:concerns}). The simplest illustrative examples are described in Section 3. Then we conclude and describe open questions and future directions. Appendices contain some complementary material and alternative derivations.

\section{Quantum mechanics}
The simplest case in which the gluing problem can be formulated, in the sense that both $\qft_d$ and $\qft_{d-1}$ exist, is that of quantum mechanics, i.e., when $d=1$. On the one hand, this is expected to be a useful theoretical laboratory where we can explicitly answer many questions which are less accessible in higher-dimensional quantum field theories. On the other hand, quantum mechanics serves as a model example for higher-dimensional problems. Indeed, if we study gluing in $d$ dimensions, then close to the boundary component $W\subset \partial M$, $M$ looks like $W\times \R$. Interpreting $\R$ as the time direction, one can think of $\qft_d$ as a quantum mechanical sigma model with the infinite dimensional target (configuration space) given by the space of fields on $W$, denoted $\text{Fields}[W]$. From this point of view, passing from quantum mechanics to higher-dimensional quantum field theories corresponds to passing from finite-dimensional to infinite-dimensional configuration spaces. We know from our experience with quantum field theories that finite-dimensional models often capture the essential phenomena. The only new things that come with increased ``infinite-dimensionality'' are regularization and renormalization, as well as a possibility of new anomalies. As we will argue later, under the assumption that $\qft_d$ is anomaly-free and properly renormalized, the associated gluing theory $\qft_{d-1}$ is automatically well-defined. Therefore, we expect to be able to capture all essential features of gluing already at the level of quantum mechanics.

In what follows, we will briefly review some relevant notions from geometric quantization (see \cite{Woodhouse:1992de} for a detailed review), and then proceed to describe boundary conditions, boundary terms, and gluing in the path integral formulation of quantum mechanics. The reason that we have to work with path integrals is that this is an essential ingredient in the physical understanding of higher-dimensional QFTs, even though in $d=1$ it could be completely bypassed. After that we will comment on how gluing works in gauge theories, i.e., constrained Hamiltonian systems with the first class constraints. Finally, we will comment on the analytic continuation relevant for connecting the Lorentzian and Euclidean time wave functions: while trivial in the ``position representation'', it becomes important when we work with more general polarizations, such as the ``momentum representation''.

\subsection{Polarization and geometric quantization}
Quantum mechanical systems are often associated with quantization of some $2m$-dimensional symplectic manifold $(X,\omega)$. A quantization produces (in a non-canonical way) from $(X, \omega)$ an algebra of quantum observables, which is a non-commutative deformation of $C^\infty(X)$, as well as its representation on a topological vector space $\cH$ (``the Hilbert space'') equipped with the appropriate Hermitian inner product. One often picks a special observable $H$---the Hamiltonian---which generates temporal evolution.

As an additional layer of structure, one can associate a special state $\psi_L\in\cH$ to each Lagrangian submanifold of $(X,\omega)$. Lagrangian submanifolds are locally defined by $m=\frac12\dim X$ independent equations that Poisson-commute with each other (with respect to Poisson brackets associated to $\omega$), and so at the quantum level they correspond to a common eigenstate of a maximal set of mutually commuting operators. This common eigenstate is $\psi_L$. (We do not discuss its properties any further.)

All these ideas can be made explicit in geometric quantization, which provides a concrete realization of the quantization procedure. Since there is no canonical quantization, any possible construction should take some arbitrary choice as an input data. For geometric quantization, the main such choice is a polarization of $(X, \omega)$.

Any symplectic manifold can be locally written, in the Darboux coordinates, as $\R^{2m}$ with coordinates $(q^i, p_i)$, $i=1\dots m$, and the symplectic form $\omega=\sum_i \dd p_i\wedge \dd q^i$. When we can do this globally, we are in the simplest situation of canonical quantization, where states are described as square-integrable functions of $q^i$ only, and quantum observables are constructed from $q^i$ and $p_i$, where the latter acts as $-i\partial/\partial q^i$ on states. The wave functions depending only on $q^i$ can be characterized as living in the kernel of $\partial/\partial p_i$, $i=1\dots m$. When it is not possible to choose Darboux coordinates globally, or we do not wish to do so, one has to proceed differently. First of all, instead of considering functions, one generally studies sections of a certain line bundle $L$ on $X$ (a prequantization bundle). Second, one equips this bundle with a connection $\nabla$ (whose curvature is $\omega/2\pi$) and demands that all sections representing physical states are annihilated by $\nabla_\cP$, where $\cP$ is a set of vectors analogous to $\partial/\partial p_i$. One has to choose such vectors in tangent spaces to every point of $X$. They have to satisfy a certain consistency condition, meaning that such vectors at the adjacent points of $X$ are not completely independent. This choice of $\cP$ is called polarization, and the precise definition is as follows. A complex polarization is a choice of a complex distribution $\cP\subset T_\C X = TX\otimes \C$ which satisfies the following conditions:\footnote{A distribution of dimension $n$ on $X$ is an $n$-dimensional subbundle of the tangent bundle $TX$. Complex distribution is a subbundle of $T_\C X= TX\otimes \C$.} 
\begin{itemize}
	\item For each $p\in X$, $\cP_p\subset T_p X \otimes \C$ is Lagrangian, i.e. $\omega(u,v)=0$, $\forall u,v\in \cP_p$ and $\dim_\C \cP=m=\frac12 \dim X$.
	\item $\cP$ is integrable, which means $[\cP,\cP]\subset \cP$, where $[\cdot, \cdot]$ denotes the commutator of vector fields.
	\item $\cP\cap \bar{\cP}$ has constant dimension throughout $X$, where $\bar{\cP}$ is a complex conjugate of $\cP$.
\end{itemize}
In fact, this $\cP\cap \bar{\cP}=D_\C$ is a complexification of a certain real distribution $D\subset TX$ (and $\cP+\bar{\cP}=E_\C$ is a complexification of another real distribution usually denoted $E\subset TX$). Integrability of $\cP$ implies integrability of $D$. Then $D$, by Frobenius theorem, can be integrated to define a foliation of $X$ by leaves of this distribution -- submanifolds whose tangent spaces coincide with $D$ at every point. These leaves are isotropic submanifolds, meaning that symplectic form $\omega$ vanishes when restricted to any leaf. The space of such leaves is denoted $X/D$.

The geometric quantization construction describes physical states as half-densities on $X/D$ with values in the prequantization bundle. One important particular case is the real polarization, which is defined by demanding that:
\begin{equation}
\cP = \bar{\cP} = \cP\cap \bar{\cP} = D_\C.
\end{equation}
In this situation, $D$ is a middle-dimensional real distribution, and by integrating it we get a foliation of $X$ by Lagrangian submanifolds. If we denote coordinates on the space of leaves $X/D$ by $x$, we can formally consider states $|x\rangle$ labeled by Lagrangian submanifolds (only formally, because such states are not square-integrable in general, and should be understood in terms of projection operators $\int_{x\in \text{some region}} \dd x\, |x\rangle\langle x|$, as is always the case for continuous spectra). Then any wave function $\psi(x)$ is formally thought of as $\langle x|\psi\rangle$ with $\psi\in\cH$, and we have the unitary evolution kernel:
\begin{equation}
\label{transition}
U(x_1, x_2; t)=\langle x_1| e^{-i tH}|x_2\rangle.
\end{equation}
As we will discuss in the next subsection, such Lagrangian submanifolds labeled by $x$ should be used as boundary conditions in the path integral formulation.

Another useful polarization is the holomorphic polarization: the distribution $\cP$ is generated by all (anti-)holomorphic vector fields on $X$; it exists and plays important role when $X$ is K\"ahler, and it satisfies:
\begin{equation}
\cP\cap \bar{\cP} = 0.
\end{equation}
Since $D=0$, in this case $X/D=X$ itself. The holomorphic polarization played a remarkable role in the study of Chern-Simons theory and 2d CFT \cite{Witten:1988hf,Axelrod:1989xt,Verlinde:1989ua}.

One can also consider more general complex polarizations, such as when $D^\C=\cP\cap \bar{\cP}$ is non-trivial but not maximal, i.e., leaves of $D$ are isotropic but not Lagrangian submanifolds. In this paper we are not going to study these, neither will we consider holomorphic polarizations. Our main focus will be on real polarizations, even though we will make some comments on complex polarization at the end. It would be interesting to understand how to extend the gluing formalism of this paper to the case of complex polarizations. To be more precise, it is of course possible to glue using complex polarizations (one has to be careful with analytic continuation and integration cycles in the path integral); what is not clear, though, is whether such gluing can preserve symmetries of $\qft_d$ in a manner similar to how gluing based on real polarization does in the current paper.

\subsection{Path integral description and polarized boundary conditions}\label{sec:polbndry}
Our main task in the present paper is to understand gluing of quantum fields theories in the path integral formalism. It is natural to address this sort of questions using the phase space path integral. The reason is that in the phase space description, equations of motion involve only first order derivatives, hence boundary conditions do not involve any derivatives at all and are simply given by submanifolds of the phase space $X$.

\begin{figure}[t!]
	\centering
	\includegraphics[width=0.5\textwidth]{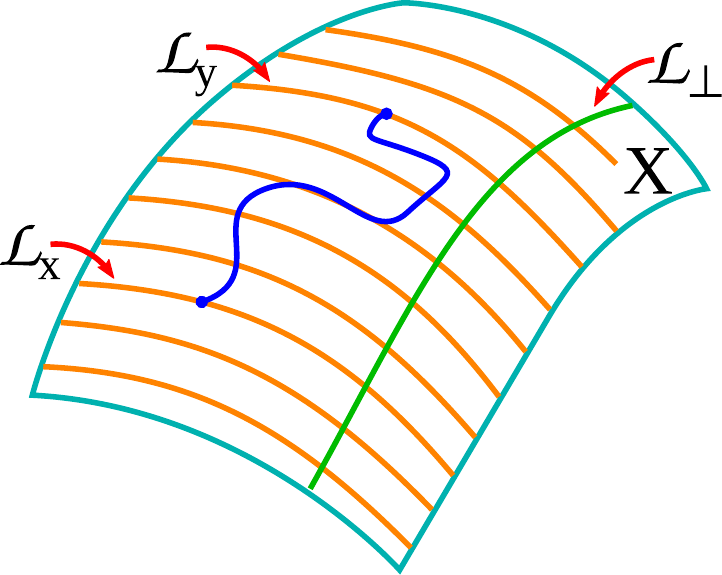}
	\caption{\label{fig:fol} Orange curves represent foliation of $X$ by Lagrangian submanifolds corresponding to a given real polarization $\cP$. Blue curve is a trajectory starting at $\cL_x$ and ending at $\cL_y$; we integrate over such trajectories. Green curve represents a Lagrangian submanifold $\cL_\perp$ that is transverse to the polarization; it might be convenient to think of physical states written in polarization $\cP$ as half-densities on $\cL_\perp$.}
\end{figure}

It is well-known that boundary conditions that have a good quantum interpretation are given by Lagrangian submanifolds of $X$. This is also where one makes contact with the formalism reviewed in the previous subsection. Namely, suppose we fix some real polarization $\cP=D_\C$ of $X$ and integrate it to a foliation of $X$ by Lagrangian submanifolds. The space of leaves $X/D$ is the space of ``position variables'', and suppose we pick two points $x, y\in X/D$. They are represented by Lagrangian submanifolds $\cL_x$ and $\cL_y$. In the phase space path integral, in order to evaluate the transition amplitude like \eqref{transition} between the points $x$ and $y$, we integrate over all paths on $X$ that start somewhere on $\cL_x$ and end on $\cL_y$, as shown in the Figure \ref{fig:fol}.

In order to have a slightly more concrete description of the quantum states, instead of working with the quotient space $X/D$, we could pick another Lagrangian submanifold, call it $\cL_\perp$, transversal to the polarization. Points of this submanifold represent equivalence classes of the quotient $X/D$, and quantum states can be described as half-densities on $\cL_\perp$. (See Figure \ref{fig:fol}.) Indeed, in geometric quantization, physical states are covariantly constant along the leaves of $D$, and so they are uniquely determined by their restriction to $\cL_\perp$.

To write the action in the phase space path integral, we proceed as follows. First, note that locally it is possible to pick such Darboux coordinates $(q^i, p_i)$ that polarization is generated by the vector fields $\partial/\partial p_i$. Then leaves of $D$ have constant $q^i$ and are parametrized by $p_i$. The submanifold $\cL_\perp$ in such coordinates is then determined by $p_i = W_i(q)$. Demanding that it is Lagrangian actually implies that $W_i(q)=\partial_i W(q)$, i.e., we have:
\begin{equation}
\cL_\perp = \left\{(q,p): p_i=\frac{\partial W(q)}{\partial q^i}\right\}.
\end{equation}
We can further define another coordinate system:
\begin{equation}
Q^i=q^i,\quad P_i = p_i - \frac{\partial W(q)}{\partial q^i}.
\end{equation}
These are also Darboux coordinates, meaning that $\omega=\sum_i \dd P_i\wedge \dd Q^i$. Furthermore, the polarization is still generated by the vector fields $\partial/\partial P_i$. One simplification is that now the manifold $\cL_\perp$ is given by:
\begin{equation}
\cL_\perp = \left\{(Q,P): P_i=0\right\}.
\end{equation}
When working in such a coordinate system, the phase space action (in Minkowski time) takes the well-known form:
\begin{align}
\label{phasePI}
S&=\int P_i \dd Q^i - H \dd t.
\end{align}
The first term here is the integral of a local one-form:
\begin{equation}
\theta = P_i \dd Q^i,
\end{equation}
which is known as the symplectic potential, and it also plays a role of the local connection one-form on the prequantization line bundle in the geometric quantization formalism. Therefore, the first term in the phase space action is almost universal and is given by:
\begin{equation}
\label{sympl_term}
\int_\gamma\theta,
\end{equation}
where $\gamma\subset X$ is the trajectory in the phase space. Here we are ignoring any possible issues related to the global topology of $X$, and essentially assuming that $\theta$ can be written globally. This is the case in all applications discussed later in this paper, however, it would be interesting to understand global properties too. In all examples we study, the phase space $X$ is a total space of the cotangent bundle $T^*Y$ to the configuration space $Y$, and so $\theta$ is (cohomologous to) the canonical one-form of $T^*Y$.

The reason we called \eqref{sympl_term} ``almost universal'' is that this term is not completely unique. The main property of $\theta$ is that locally $\dd\theta=\omega$, and this property is preserved under $\theta \to \theta + \dd F$, which are gauge transformations of the prequantization bundle. If $\gamma: [0,1] \to X$, this freedom gives rise to:
\begin{equation}
\int_\gamma\theta \to \int_\gamma \theta + F\big|^{\gamma(1)}_{\gamma(0)}.
\end{equation}
However, in the action \eqref{phasePI}, there is no such freedom, and for a very good reason: adding ``boundary terms'' $F\big|^{\gamma(1)}_{\gamma(0)}$ at the endpoints of $\gamma$ corresponds to changing the allowed boundary conditions in the phase space path integral. Let us understand this more precisely.

Already from the way the path integral formulation is usually ``derived'' from the operator approach (by inserting the decomposition of identity multiple times, either in the form $1=\int |Q\rangle \dd Q \langle Q|$ or $1=\int |P\rangle \dd P \langle P|$), it is clear that the action \eqref{phasePI} assumes boundary conditions fixing $Q$ at the endpoints of $\gamma$. It is also known (for example from the Fourier transform) that if we wish to fix $P_i$ at the endpoints, we should write the action in a different form, namely:
\begin{equation}
\label{fixP}
S=\int_\gamma (-Q^i \dd P_i - H \dd t) = -Q^iP_i\big|^{\gamma(1)}_{\gamma(0)}   + \int (P_i \dd Q^i - H \dd t),
\end{equation}
which differs from \eqref{phasePI} by the boundary term $-P_i Q^i\big|^{\gamma(1)}_{\gamma(0)}$.
More general boundary conditions given by more general Lagrangian submanifolds give rise to more general boundary terms.

One possible way to understand the need for boundary terms is by looking at the equations of motion. In order for the path integral to have a good saddle-point approximation, the boundary conditions should be picked in such a way that classical equations of motions admit a unique solution. Then expanding around the classical trajectory provides a well-defined saddle point approximation. Taking variation of the Minkowski signature action \eqref{phasePI} gives:
\begin{equation}
\delta S = P_i \delta Q^i\big|^{\gamma(1)}_{\gamma(0)} + \int \left[\delta P_i (\dot{Q}^i - \partial H/\partial P_i) - \delta Q^i (\dot{P}_i + \partial H/\partial Q^i) \right] \dd t.
\end{equation}
We see that this yields standard Hamiltonian equations of motion only under the assumption that $P_i \delta Q^i$ vanishes at the endpoints. For this to vanish, one should either have $P_i=0$ or $\delta Q^i=0$. The former gives an isolated boundary condition that we are not interested in. The latter, on the other hand, gives a family of boundary conditions parametrized by the fixed value of $Q^i\big|$ at the boundary, and this is the boundary condition which allows to calculate the transition amplitude $\langle Q=x_1|e^{-it H}|Q=x_2\rangle$. Therefore, we say that the action written in the form \eqref{phasePI} is associated with the family of boundary conditions fixing $Q^i$ at the boundary, or the position-based polarization. In the same manner, varying \eqref{fixP} would produce a boundary variation $-Q^i\delta P_i$, which vanishes for a family of boundary conditions characterized by a fixed value of $P_i$ at the boundary. Similarly, we discard an isolated boundary condition $Q^i=0$ in this case. (Notice that these isolated boundary conditions that we discard are simply those for which the boundary term $-P_i Q^i$ from \eqref{fixP} vanishes.)

To account for more general boundary conditions given by some Lagrangian submanifolds $\cL_0$ and $\cL_1$ at the endpoints of $\gamma$, we would have to add boundary terms (or ``boundary corrections'' as we will call them later) $F_0(Q,P)$ and $F_1(Q,P)$ to the action \eqref{phasePI}, so that the variational problem becomes consistent with $\cL_0$ and $\cL_1$. This is expressed as:
\begin{align}
\label{deltaF}
(P_i \delta Q^i + \delta F_0(Q,P))\big|_{\cL_0} &= 0,\cr
(P_i \delta Q^i + \delta F_1(Q,P))\big|_{\cL_1} &= 0,\cr
\end{align}
so that the path integral:
\begin{equation}
\langle \cL_1|e^{-i H}| \cL_0\rangle = \int_{\substack{\gamma(0)\in\cL_0\\ \gamma(1)\in\cL_1}} \pD \gamma\, e^{iF_1\big|_{\gamma(1)} - iF_0\big|_{\gamma(0)} + i\int (P_i \dd Q^i  - H \dd t)}
\end{equation}
acquires a well-defined saddle-point approximation. 

However, requirements \eqref{deltaF} do not determine $F_{0,1}$ uniquely, only up to an arbitrary function that is constant along $\cL_{0,1}$. This is clear already from \eqref{phasePI}: we could add arbitrary functions of $Q$, denoted $F(Q)$, at the endpoints of $\gamma$, and because $Q$ is fixed there, it would not affect the variational problem. It corresponds to multiplying a wave function by $e^{i F(Q)}$, and is simply a gauge transformation of the prequantization bundle. Suppose that we have added such boundary terms, say $F(Q)\big|_{\gamma(1)} - F(Q)\big|_{\gamma(0)}$. We could absorb them into the symplectic potential as follows:
\begin{equation}
F(Q)\big|_{\gamma(1)} - F(Q)\big|_{\gamma(0)} + \int_{\gamma}P_i \dd Q^i = \int_{\gamma} (P_i + \partial_i F(Q))\dd Q^i.
\end{equation}
Further redefinition $P_i + \partial_i F(Q) = \tilde{P}_i$ would remove the boundary terms $F(Q)$ completely. However, the $\cL_\perp$ submanifold given by $P_i=0$ would be described by the equation:
\begin{equation}
\tilde{P}_i = \partial_i F(Q).
\end{equation}
What this discussion implies is that if we pick the submanifold $\cL_\perp$ and fix the coordinate system in such a way that $\cL_\perp$ is locally given by $P_i=0$, (in addition to the polarization being generated by $\partial/\partial P_i$,) this removes the freedom of adding arbitrary boundary terms $F(Q)$.

To be more precise, if we have chosen Darboux coordinates $(P_i, Q^i)$ such that polarization is generated by $\partial/\partial P_i$ and a transversal Lagrangian submanifold $\cL_\perp$ is determined by $P_i=0$, we still have a remaining freedom to perform a further coordinate change preserving all these conditions. Namely, we could perform a diffeomorphism:
\begin{align}
Q^i &= Q^i(x),\cr
\frac{\partial Q^j(x)}{\partial x^i} P_j &= y_i.
\end{align}
One can check that this is indeed a symplectomorphism, i.e., $\sum_i \dd P_i\wedge \dd Q^i = \sum_i \dd y_i\wedge \dd x^i$, that the polarization is now generated by $\partial/\partial y_i$, and that the submanifold $\cL_\perp$ is given by $y_i=0$. However, this symplectomorphism also preserves the canonical form of the symplectic potential:
\begin{equation}
\sum_i P_i \dd Q^i = \sum_i y_i \dd x^i.
\end{equation}
The latter means that it does not introduce any further ambiguity in the symplectic potential, which is uniquely fixed once we pick a polarization and a transversal submanifold $\cL_\perp$. Actually, this fact can be understood in a more invariant way, without using coordinates explicitly, as follows. By picking a polarization and $\cL_\perp$, we describe $X$ locally as $T^*\cL_\perp$ with the canonical symplectic structure and polarization along fibers, and any cotangent bundle has the canonical choice of the symplectic potential, which is the one we are using. The remaining freedom to perform symplectomorphisms preserving both $\cL_\perp$ and the polarization also preserves the symplectic potential.

Therefore, to summarize, if we pick a polarization $\cP$ and a transversal Lagrangian submanifold $\cL_\perp$. This allows to make a canonical choice of the symplectic potential since locally $X$ is symplectomorphic to $T^*\cL_\perp$. For that, we pick Darboux coordinates $(P_i, Q^i)$ in which $\cP$ is generated by $\partial/\partial P_i$ and $\cL_\perp$ is determined by the equations $P_i=0$, and in any such coordinate system define:
\begin{equation}
\theta_0=P_i \dd Q^i.
\end{equation}

If we start with an arbitrary coordinate system and make an arbitrary choice of $\theta$, it might differ from $\theta_0$ by a total derivative:
\begin{equation}
\theta_0=\theta + \dd F,
\end{equation}
and this $F$, evaluated at the endpoints of the phase space trajectory, must be added to the action $\int_{\gamma}(\theta - H \dd t)$ to make it consistent with the boundary conditions given by Lagrangian leaves of $\cP$. Notice that now $F$ is determined uniquely up to a constant.

To be completely general, we might be willing to write the ``in'' and ``out'' states of the path integral in different polarizations. Namely, we could be dealing with the two different polarizations $\cP_0$ and $\cP_1$, and use Lagrangian leaves of $\cP_0$ as boundary conditions at $t=0$, while Lagrangian leaves of $\cP_1$ could be taken as boundary conditions at $t=1$. This is perfectly consistent with the above discussion, the only difference being that the boundary corrections $F_0$ and $F_1$ for the two endpoints would have to be determined separately using the described procedure. In the end, we arrive at the expression:
\begin{equation}
\label{generalPI}
\langle \cL_1|e^{-i H}| \cL_0\rangle = \int_{\substack{\gamma(0)\in\cL_0\\ \gamma(1)\in\cL_1}} \pD \gamma\, e^{iF_1\big|_{\gamma(1)} - iF_0\big|_{\gamma(0)} + i\int_\gamma (\theta  - H \dd t)},
\end{equation}
where now $F_{0,1}$ are determined uniquely up to a constant. This ambiguity is harmless and corresponds to the ambiguity of multiplying wave functions by a constant phase.

To emphasize the important notion of boundary conditions given by Lagrangian leaves of a chosen real polarization, we use a special name for it:

\textbf{Definition:} For a given real polarization $P=D_\C$, boundary conditions determined by Lagrangian integral submanifolds for $D$ are called polarized boundary conditions.

\subsubsection{A remark on choices}
One way to understand the need to pick $\cL_\perp$ in order to write an unambiguous phase space path integral is as follows. When we pass from the operator formalism (say, developed based on the geometric quantization techniques,) to the path integral description, we usually discretize time and insert identity in the form $\int |q\rangle \dd q \langle q|$ and $\int |p\rangle \dd p \langle p|$ multiple times. To do this, we actually have to choose two complementary polarizations, one called $\cP$, with the leaves labeled by $q$, and another called $\cP_\perp$, with the leaves labeled by $p$. The quantum states formally associated to Lagrangian leaves of $\cP$ are given by $|q\rangle$ (modulo usual technicalities associated with the continuum spectrum), while those for $\cP_\perp$ are called $|p\rangle$. Then $\cL_\perp$ is simply one of the leaves of $\cP_\perp$, the most convenient choice being $p=0$.

\subsection{Gluing and symmetry}\label{sec:gluesym}
The main lesson of Subsection \ref{sec:polbndry} is that, in general, if space-time has a boundary, then boundary terms have to be included in the path integral, like in \eqref{generalPI}. Moreover, the precise form of boundary terms is determined by the choice of polarization used for the description of quantum states. Boundary terms also become important once we start to act with symmetries, as we will see soon.

Let us now address the question of gluing. Suppose we have a state vector $|\psi_1\rangle$ and a covector $\langle\psi_2|$. We might think of $|\psi_1\rangle$ as living at the endpoint of an interval $I_1$ and representing the result of some quantum dynamics taking place inside the interval. Similarly $\langle\psi_2|$ lives at the endpoint of $I_2$. We think of $|\psi_1\rangle$ as living on a positively oriented endpoint (the ``output'') of the first interval, and $\langle \psi_2|$ as living on a negatively oriented endpoint (the ``input'') of the second interval. As we know, we can glue the two intervals together by simply computing $\langle\psi_2|\psi_1\rangle$. To write this sewing procedure explicitly, we pick a real polarization on $X$, choose a coordinate $x$ on its space of leaves $X/D$ (which we identify with $\cL_\perp$), and describe both states in terms of this polarization as $\langle x|\psi_1\rangle$ and $\langle \psi_2|x\rangle$:
\begin{equation}
\label{sewing}
\langle \psi_2|\psi_1\rangle = \int_{\cL_\perp} \dd x \langle \psi_2|x\rangle\langle x|\psi_1\rangle.
\end{equation}

This is the most standard quantum-mechanical fact, except that we choose to formulate it in terms of an arbitrary polarization. Now let us address the following question: when does a symmetry of the original theory induce a symmetry of the integral in \eqref{sewing} (which can be considered as a zero-dimensional QFT)? At the very least, this symmetry must \emph{act} on \eqref{sewing}, meaning that it should transform $x$ variables into $x$ variables, without mixing them with other directions of the phase space. Since $x$ parametrizes the space of leaves $X/D$, this means that the symmetry should act on this space, transforming one Lagrangian leaf into another. The latter means that it transforms any tangent space to a Lagrangian leaf to another such space, which can be simply reformulated as a condition that the symmetry preserves polarization. If our symmetry is generated by $\Phi\in C^\infty(X)$, and the corresponding vector field is:
\begin{equation}
V_\Phi = \omega^{-1}(d\Phi, \cdot),
\end{equation}
where $\omega^{-1}$ is the Poisson structure, then the condition that it preserves polarization $\cP$ is:
\begin{equation}
[V_\Phi,\cP]\subset \cP,
\end{equation}
where $[\cdot,\cdot]$ is the commutator of vector fields. If we pick, as usual, local Darboux coordinates $(P_i, Q^i)$ in which $\cP$ is generated by $\partial/\partial P_i$, then the generator of the most general symmetry that preserves $\cP$ can be written in such coordinates as:
\begin{equation}
\label{Sym_gen}
\Phi=g(Q)+\sum_i P_i f^i(Q),
\end{equation}
which corresponds to the vector field:
\begin{equation}
\label{Sym_vect}
V_\Phi = -f^i(Q) \frac{\partial}{\partial Q^i} + \left( \frac{\partial g(Q)}{\partial Q^i} + P_j \frac{\partial f^j(Q)}{\partial Q^i} \right)\frac{\partial}{\partial P_i}.
\end{equation}
This follows immediately from the fact that the coefficient of $\partial/\partial Q^i$ in $V_\Phi$ equals $-\partial\Phi/\partial P_i$, and it should be some $P_i$-independent function $f^i(Q)$ in order for $[V_\Phi, \cP]\subset \cP$ to hold. This then integrates to \eqref{Sym_gen}.

Now let us understand how this symmetry acts on the gluing formula \eqref{sewing}. It is somewhat instructive to see how it works in the path integral formalism, however the related discussion is slightly technical, and for this reason it is presented in Appendix \ref{appen:main}. Here we provide a shortcut argument. We are going to claim that $\Phi$ becomes a symmetry of \eqref{sewing} if the states $\psi_1$ and $\psi_2$ are annihilated by the corresponding quantum generator $\hat\Phi$. Let us resolve ordering ambiguity in a way that makes $\hat\Phi$ Hermitian (final conclusion does not depend on this choice):
\begin{equation}
\hat\Phi= g(Q) + \frac12 \sum_i(\hat{P}_if^i(Q) + f^i(Q)\hat{P}_i),
\end{equation} 
where $\hat{P}_i$ acts by $-i\partial/\partial Q^i\equiv -i\partial_i$. We also identify $x=Q$, where $x$ is the coordinate on the space of leaves $X/D \cong \cL_\perp$. We have:
\begin{align}
0=\langle x|\hat\Phi |\psi_1\rangle = \left( g(x) - \frac{i}2 \partial_i f^i(x) - i f^i(x)\partial_i\right)\langle x|\psi_1\rangle,
\end{align}
or, by writing $(1+i\epsilon\hat\Phi)|\psi_1\rangle=|\psi_1\rangle$ and assuming $\epsilon$ is infinitesimal, this can be expressed as:
\begin{equation}
\langle x+\epsilon f(x)|\psi_1\rangle=e^{-i\epsilon g(x)-\frac{\epsilon}{2} \partial_i f^i(x)}\langle x|\psi_1\rangle.
\end{equation}
Analogously, when acting on the left:
\begin{align}
0=\langle \psi_2|\hat\Phi |x\rangle = \left( g(x) + \frac{i}2 \partial_i f^i(x) + i f^i(x)\partial_i\right)\langle \psi_2|x\rangle,
\end{align}
which is written infinitesimally as:
\begin{equation}
\langle\psi_2|x+\epsilon f(x)\rangle = e^{i\epsilon g(x) - \frac{\epsilon}{2}\partial_i f^i(x)}\langle\psi_2|x\rangle.
\end{equation}
Altogether, this implies:
\begin{equation}
\label{dens_tr}
\langle\psi_2|x+\epsilon f(x)\rangle\langle x+\epsilon f(x)|\psi_1\rangle =e^{-\epsilon\partial_i f^i(x)} \langle\psi_2|x\rangle\langle x|\psi_1\rangle.
\end{equation}
Now let us perform a diffeomorphism
\begin{equation}
\label{symm0d}
x^i \to \tilde{x}^i + \epsilon f^i(\tilde{x})
\end{equation}
in the gluing integral \eqref{sewing}. The measure $\dd x$ will produce a Jacobian factor $1 + \epsilon \partial_i f^i(\tilde{x})=e^{\epsilon \partial_i f^i(\tilde{x})}$ ($\epsilon$ is still assumed to be infinitesimal). Then we obtain:
\begin{align}
\langle \psi_2|\psi_1\rangle = \int_{\cL_\perp} \dd \tilde{x}\, e^{\epsilon \partial_if^i(\tilde{x})}\left\langle\psi_2|\tilde{x}+\epsilon f(\tilde{x})\right\rangle\left\langle \tilde{x}+\epsilon f(\tilde{x})|\psi_1\right\rangle = \int_{\cL_\perp} \dd \tilde{x} \langle\psi_2|\tilde{x} \rangle\langle \tilde{x}|\psi_1\rangle,
\end{align}
where in the last equality we used \eqref{dens_tr}. This line of reasoning proves:

\textbf{The Main Lemma:} The gluing integral \eqref{sewing}, considered as a zero-dimensional QFT, has a symmetry \eqref{symm0d} provided that the states $\psi_1$ and $\psi_2$ are $\hat\Phi$-invariant, and $V_\Phi$ preserves polarization $\cP$ on the phase space.

\subsubsection{Anomaly inflow}
A slightly more accurate way to prove the above Lemma would be to consider half-densities $\rho(\psi_1)=\langle x|\psi_1\rangle \sqrt{\dd x}$ and $\rho(\psi_2)=\langle\psi_2|x\rangle\sqrt{\dd x}$, rather than separate $dx$ from the wave-functions. The reason is that states are more naturally described by half-densities, as we know from the geometric quantization formalism. Such a point of view would also eliminate the logical step where we had to ``cancel'' the Jacobian factor $e^{\epsilon \partial_i f^i(x)}$ against the similar one coming from the integrand $\langle\psi_2|x\rangle\langle x|\psi_1\rangle$. Moreover, only the total density $\rho(\psi_2)\rho(\psi_1)$ has a chance of being generalizable to higher-dimensions, where the finite integral \eqref{sewing} would be replaced by a path integral, whereas generalizing $dx$ separately to the ``measure'' $\pD \phi$ is usually an ill defined step.

Nevertheless, there is a certain value in proving the Lemma in the way we did, even though it was slightly ``unnatural''. The reason is that it provides a simple illustration of the anomaly inflow mechanism. Indeed, if we wanted to consider the measure $\dd x$ separately and use it to build a zero-dimensional QFT with some action $S(x)$ invariant under \eqref{symm0d}, such a theory would not posses this symmetry in general, simply because the measure $\dd x$ is not invariant, unless $\partial_i f^i(x)$ vanishes. However, in the gluing theory \eqref{sewing}, the bulk contribution $\langle \psi_2|x\rangle\langle x|\psi_1\rangle$ is similarly non-invariant and cancels the $\partial_i f^i(x)$ term coming from the measure. In this sense, it can be regarded as an example of the anomaly inflow.

The way it works in higher dimensions is exactly the same: once we know that the $d$-dimensional theory is non-anomalous, so is the $(d-1)$-dimensional gluing theory, even if the field content and symmetries look like it could be anomalous. One example will be gluing of 3D gauge theories with fermionic matter using the chiral polarization discussed in Section \ref{sec:anomglue}: the gluing theory will be a 2D gauge theory with chiral fermions. Such a theory would usually be inconsistent due to gauge anomalies, however it makes perfect sense as a gluing theory thanks to the anomaly inflow from the bulk.

Furthermore, in Section \ref{sec:globanom} we will give a simple example of gluing where the gluing theory appears to have a $\Z_2$ global anomaly. Nevertheless, consistency of the parent theory implies that the global anomaly should be canceled by a similar $\Z_2$ non-invariance of the bulk contribution. Strictly speaking, this type of anomaly is not covered by the above quantum-mechanical derivation, which only applies to local anomalies. Nevertheless, we still claim that it should cancel, simply because the parent theory is consistent and local, and thus the gluing property should hold.

\subsection{Gauge theories}\label{gauge_general}
Discussion in the previous section is quite general: even though we ignore global issues related to possible non-trivial topology of the phase space, it always applies locally, and the final conclusion about gluing should hold generally. In particular, it applies to gauge theories if we properly perform gauge fixing first. 

The modern way to do this at the highest level of generality is by applying the BV formalism: introducing additional fields and modifying the action, one trades gauge symmetry for the global odd BRST symmetry generated by $Q_B$. More precisely, since we are interested in manifolds with boundary, one should use the BV-BFV formalism of \cite{Cattaneo:2012qu,Cattaneo:2012zs,Alekseev:2012wc,Cattaneo:2015vsa,Cattaneo:2016hqk,Cattaneo:2017tef,Iraso:2018mwa} that was already mentioned in the introduction. In this case, $Q_B$ corresponds to an odd vector field in the extended phase space. Since it generates the global symmetry, we can simply apply our previous results and determine when it induces symmetry in the gluing theory. Every physical state is $Q_B$-closed, hence if the polarization is also $Q_B$-invariant, the gluing theory acquires an induced odd symmetry $\tilde{Q}_B$. This $\tilde{Q}_B$ is the BV-BRST charge of the gluing theory, hence the gluing theory must be a gauge theory as well.

Unfortunately, this argument is not constructive. Furthermore, we are not going to discuss the BV-BFV formalism here -- interested readers should consult the above references. Instead, much in the spirit of the rest of this paper, we take a more simply-minded approach, which makes sense in finite-dimensional cases. It will provide us with a simple and intuitive geometric picture for why the gluing theory is a gauge theory. Instead of fixing gauge, we describe gluing in a gauge-covariant way, such that the gluing theory has a manifest gauge symmetry. 

From the canonical formalism point of view, gauge theories are constrained Hamiltonian systems with the first class constraints.\footnote{If there are also second class constraints present, we may assume that they have been resolved, and the induced Poisson bracket on the constraint surface, known as the Dirac bracket, has been constructed.} Suppose that we start with a symplectic manifold $(N, \omega)$---the total phase space---and have a set of first class constraints defined on it:
\begin{equation}
\phi_i=0,\, i=1\dots r.
\end{equation}
The fact that they are first class is usually expressed as:
\begin{equation}
\label{firstCL}
\{\phi_i, \phi_j\} = C_{ij}^k \phi_k,
\end{equation}
where $\{\cdot,\cdot\}$ is a Poisson bracket, and $C_{ij}^k\in C^\infty(N)$. The constraint surface is:
\begin{equation}
\{x\in N|\, \phi_i(x)=0\} = C \subset N.
\end{equation}
Then, as a consequence of \eqref{firstCL}:
\begin{equation}
\{\phi_i,\phi_j\}\big|_C = 0.
\end{equation}
Let $V_i$ denote Hamiltonian vector fields associated to $\phi_i$, i.e., $\dd\phi_i + \iota_{V_i}\omega=0$. The above equation means that on $C$, they are tangent to $C$. These vector fields on $C$, equipped with their commutator as a Lie bracket, form a Lie algebra $\mathfrak{g}$ -- the algebra of gauge transformations, which exponentiates to the group of gauge transformations\footnote{Not to confuse with the gauge group. In higher dimensional gauge theories on $W\times \R$ with the gauge group $G$, this $\cG$ would be ${\rm Hom}(W, G)$.} $\cG$. By taking quotient, we get the physical phase space of the theory:
\begin{equation}
X = C/\cG.
\end{equation}
This is nothing else but the standard symplectic reduction, with constraints $\phi_i$ defining the moment map. The manifold $X$ inherits symplectic structure in a canonical way (due to the Marsden-Weinstein theorem). The pullback of this symplectic structure from $X$ to $C$ with respect to the canonical projection,
\begin{equation}
{\rm pr}: C \rightarrow C/\cG,
\end{equation}
coincides with the symplectic structure of $N$ restricted to $C$, i.e., $\omega|_C$. This $\omega|_C$ is a closed two-form, but it is degenerate on $C$ of course: all $V_i$ span its kernel.

Boundary conditions for the theory with the phase space $X$ are given, as usual, by Lagrangian submanifolds $\cL\subset X$. These submanifolds lift to the maximal isotropic submanifolds $\tilde{\cL}$ of $C$ (with respect to $\omega|_C$), and under the further embedding $\tilde{\cL}\subset C\subset N$, they are Lagrangian submanifolds of $N$ invariant under the gauge group action.

Suppose we have fixed a real polarization $\cP$ on $X$ and are studying the path integral for this theory that produces some boundary state written in terms of this polarization. According to the formalism described in earlier sections, for a given choice of symplectic potential $\theta$, there is a canonical choice of the boundary term (up to a constant), and the path integral is given, say, by:
\begin{equation}
\int_{x(0)\in \cL} \pD x\, e^{i F(0) + i\int \theta - \int H \dd t},
\end{equation}
where only at $t=0$ we have explicitly specified a polarized boundary condition given by some Lagrangian submanifold $\cL$ coinciding with one of the leaves of polarization $\cP$, and $F(0)$ is the corresponding boundary term.

Now we can pull $\theta$ back to $C$, i.e., consider ${\rm pr}^*\theta$ instead of $\theta$. $H$, being a function on $X = C/\cG$, lifts to a $\cG$-invariant function on $C$. Therefore, in the above path integral, instead of integrating over trajectories in $X$, we could integrate over trajectories in $C$. This, of course, introduces a lot of unphysical freedom: every trajectory can be wiggled arbitrarily in the direction of $\cG$ orbits without changing the action. In other words, if we integrate over trajectories into $C$, the above phase space path integral has gauge redundancies. This is the original gauge theory in a ``gauge-unfixed'' form. One has to fix gauge redundancies, and of course we know how to do this: the non-redundant, i.e. gauge-fixed, path integral goes over trajectories into $X$, not $C$.

What this discussion gives us is the following. The path integral over trajectories into $X$ produces the boundary state $\psi$ that can be described as a half-density $\rho(\psi)$ (or section-valued half-density) on some Lagrangian submanifold $\cL_\perp$, transversal to the polarization $\cP$ on $X$. Instead, we could take a gauge-unfixed perspective, and extend this Lagrangian $\cL_\perp$ to a maximal isotropic submanifold $\widetilde{\cL}_\perp\subset C$. Then the boundary state generated by the path integral would extend to a $\cG$-invariant half-density $\rho(\widetilde{\psi})$ on this $\widetilde{\cL}_\perp$.

As we glue states, we could either use $\psi$ or $\widetilde\psi$. In the former case, we simply write:
\begin{equation}
\label{gaugeglue1}
\int_{\cL_\perp} \rho(\psi_1) \rho(\psi_2),
\end{equation}
and this is obviously correct as it originates from the quantization of the physical phase space. In the latter case, when we use $\tilde\psi$, the gluing could be naively expressed as:
\begin{equation}
\label{gaugeglue2}
\int_{\widetilde{\cL}_\perp} \rho(\widetilde{\psi}_1) \rho(\widetilde{\psi}),
\end{equation}
but this integral has a gauge symmetry that has to be further fixed: $\cG$ acts on $\widetilde{\cL}_\perp$, and the integrand is $\cG$-invariant. Recalling that $\cG$ is the group of gauge transformations generated by constraints, it is clear how to modify \eqref{gaugeglue2} to put it in agreement with \eqref{gaugeglue1}, at least formally:
\begin{equation}
\int_{\widetilde{\cL}_\perp} \frac{\rho(\widetilde{\psi}_1) \rho(\widetilde{\psi})}{{\rm Vol}(\cG)} = \int_{\cL_\perp} \rho(\psi_1) \rho(\psi_2).
\end{equation}

To recapitulate, this is useful in gauge theories, where the boundary states are gauge-invariant due to the Gauss law constraint, and we find that the gluing theory itself has a gauge symmetry\footnote{For completeness, we should note that the boundary wave function, even though annihilated by the Gauss law constraint, may not be gauge-invariant as a functional if there is a non-trivial anomaly inflow happening. However, the gluing theory is always gauge-invariant.}. If the theory is a pure gauge theory and we use Dirichlet polarization, our prescription reduces to the following recipe found in the literature: ``Dirichlet boundary conditions break gauge symmetry at the boundary, leaving the leftover global symmetry $G$ there; gluing can be performed via gauging the diagonal of $G\times G$ by adding the gauge multiplets that couple dynamics on the two sides of the gluing surface.'' (This procedure has been formulated in various ways, see e.g. \cite{Drukker:2010jp,Gaiotto:2014gha,Dimofte:2011py}.) Our discussion puts this into a more general perspective of cutting/gluing with symmetries, as well as provides generalizations. We will have some simple examples of the latter soon. Also, as we will learn in the applications of \cite{glue2}, we often have more than just gauge fields (or multiplets) at the gluing surface. 

Note that here we started from gluing in the gauge-fixed theory with the phase space $X$ and, by lifting from $X$ to $C$, recovered the gauge-unfixed version of the procedure. A polarization on $X$ got lifted to a gauge-invariant distribution on $C$. In all practical examples, however, we will not have the gauge-fixed version available as an input. Therefore, in order for the gluing to work, it is important to pick a gauge-invariant distribution on $C$, so that it descends to a polarization on the quotient space $X$. In all examples, it will be quite clear how to make the corresponding choice, and we will refer to this gauge-invariant distribution (describing polarization after gauge-fixing) as a polarization as well. More details will be provided as we discuss an example in Section \ref{sec:puregauge}.

\subsection{Analytic continuation and the space-time signature}\label{sec:analytic}
So far we have been working in Lorentzian time. For applications to higher-dimensional theories, it is also necessary to understand the case of Euclidean signature because a lot of higher-dimensional manifolds do not admit pseudo-Riemannian metrics. As a result of this, and because gluing procedure is most naturally formulated in Lorentzian signature, gluing of Euclidean QFTs, in general, will involve certain analytic continuation. The goal of this subsection is to explain this property at the level of quantum mechanics, which will later prove useful in our concrete applications to higher-dimensional QFTs.

Because in all higher-dimensional examples that we consider, the phase space is a cotangent bundle to another manifold, we will focus on this case for now:\footnote{This holds globally for the simplest bosonic theories based on the second order Lagrangian, in which case $Y$ is the configuration space. It also holds globally for the standard first order fermionic actions, in which case one can pick half of the fermions that determine polarization to parametrize an odd space $Y$, while the rest will define odd fibers of $T^*Y$. This statement becomes less obvious for gauge theories, but still holds for those based on the Yang-Mills action. It definitely fails globally for theories based on the first order bosonic actions such as the Chern-Simons theory, in which case the phase space is not of the form $T^*Y$ globally. Hence any global statements made specifically for $T^*Y$ should not be expected to hold immediately for the first order theories like Chern-Simons.}
\begin{equation}
X=T^*Y.
\end{equation}
Once we pick some local coordinates $q^i$ on the base $Y$, they generate a basis $dq^i$ in the fiber, and the corresponding coordinates in the fiber are denoted by $p_i$, as usual. The canonical symplectic potential on $X$ in these coordinates takes the well-known form:
\begin{equation}
\theta=p_i \dd q^i.
\end{equation}
We are going to specialize even more and consider Hamiltonians that are at most quadratic in $p_i$, again because all our examples are going to be of this sort:
\begin{equation}
H(p,q)= \frac{1}{2} g^{ij}(q)p_i p_j + a^i(q)p_i + b(q),
\end{equation}
where $g^{ij}(q)$ is a non-degenerate matrix. The standard Lorentzian time action in the Hamiltonian formalism, designed for the boundary conditions fixing $q$ at the endpoints, as we know, is:
\begin{equation}
S=\int\left(p_i \dot{q}^i - H(p,q)\right)\dd t.
\end{equation}
Integrating out $p$ gives it the on-shell value:
\begin{equation}
p_i = g_{ij}(q)(\dot{q}^j-a^j(q)),
\end{equation}
where $g_{ij}$ is the inverse matrix of $g^{ij}$. This is the textbook way to recover the Lagrangian description which reads:
\begin{equation}
S=\int\left(\frac12 g_{ij}(q)\dot{q}^i\dot{q}^j - g_{ij}(q) a^i(q) \dot{q}^j +\frac12 g_{ij}a^i(q)a^j(q) - b(q)\right)\dd t.
\end{equation}
What changes upon Wick rotation? The first order action in Euclidean time is:
\begin{equation}
S_E=\int\left(-ip_i \dot{q}^i + H(p,q)\right)\dd\tau,
\end{equation}
and the on-shell expression for $p$ acquires a factor of $i$ in front of $\dot{q}^i$:
\begin{equation}
p_i=g_{ij}(q)(i \dot{q}^j - a^j(q)),
\end{equation}
so that integrating out $p$ gives a Euclidean time second order action:
\begin{equation}
S_E=\int\left(\frac12 g_{ij}(q)\dot{q}^i\dot{q}^j + i g_{ij}(q) a^i(q) \dot{q}^j -\frac12 g_{ij}a^i(q)a^j(q) + b(q)\right)\dd\tau.
\end{equation}
We see that the reality condition of $p$ got modified on shell:
\begin{equation}
p_i + a_i(q) \to i(p_i + a_i(q)),
\end{equation}
which is totally fine at this point: we can still keep $p$ real in the phase space path integral, and it acquires a complex value only on shell. We find the usual Euclidean action in this way. However, if we wish to apply the standard saddle-point reasoning to the second order Euclidean action and expand it around the classical solution, we encounter a small problem: equations of motion following from this action are complex because of ``$i$'' in front of the second term in the action (coupling to the magnetic potential $a_i(q)$). In principle, this is not terrible, saddle points in the complex Gaussian integrals often turn out to be outside the integration cycle, but we can do slightly better by analytically continuing the magnetic field. Namely, consider a slightly modified Euclidean action:
\begin{equation}
S_E(s)=\int\left(\frac12 g_{ij}(q)\dot{q}^i\dot{q}^j + s g_{ij}(q) a^i(q) \dot{q}^j -\frac12 g_{ij}a^i(q)a^j(q) + b(q)\right)\dd\tau,
\end{equation}
where $s$ is now a real parameter. This action gives real EOMs with some real solution, and we can, in principle, expand around such a real saddle and compute the path integral. In the answer, we then have to go back to imaginary $s$:
\begin{equation}
s \to i,
\end{equation}
in order to recover the correct answer for the initial Euclidean action $S_E=S_E(s=i)$. This might seem as an unnecessary complication, but it will in fact be helpful in a moment.

Let us now see what happens if we choose to start from the first order action with the boundary terms included. For concreteness, we include boundary terms that make it consistent to fix $p_i$ at the endpoints. The corresponding Lorentzian time first order action is: 
\begin{equation}
S=-p_iq^i\big|_0^T + \int_0^T\left(p_i \dot{q}^i - H(p,q)\right)\dd t,
\end{equation}
and we impose $p(0)=p^{(1)}$, $p(T)=p^{(2)}$. Integrating out $p$ produces a second order action with the boundary terms:
\begin{equation}
S=p^{(1)}_i q^i(0)- p^{(2)}_i q^i(T) +\int\left(\frac12 g_{ij}(q)\dot{q}^i\dot{q}^j - g_{ij}(q) a^i(q) \dot{q}^j +\frac12 g_{ij}a^i(q)a^j(q) - b(q)\right)\dd t.
\end{equation}
It is easy to check, by looking at the boundary contribution in the equations of motion, that this action is consistent with the boundary conditions:
\begin{align}
\dot{q}_i(0) - a_i(q(0)) &= p_i^{(1)},\cr
\dot{q}_i(T) - a_i(q(T)) &= p_i^{(2)}.
\end{align}
This illustrates that the general boundary conditions in the phase space path integral with corresponding boundary corrections included have their counterpart in the configuration space path integral, with the corresponding boundary corrections included.

Let us see how this works in Euclidean time. Wick rotating the phase space action gives:
\begin{equation}
S_E=ip_iq^i\big|_0^T + \int_0^T\left(-ip_i \dot{q}^i + H(p,q)\right)\dd\tau.
\end{equation}
After integrating out $p$, we obtain the configuration space Euclidean action:
\begin{equation}
S_E=-ip^{(1)}_i q^i(0)+ ip^{(2)}_i q^i(T) +\int\left(\frac12 g_{ij}(q)\dot{q}^i\dot{q}^j +i g_{ij}(q) a^i(q) \dot{q}^j -\frac12 g_{ij}a^i(q)a^j(q) + b(q)\right)\dd\tau.
\end{equation}
If we look at the boundary conditions that are induced by the boundary terms in this action, they take the form:
\begin{align}
\dot{q}_i(0) + i a_i(q(0)) &= -i p_i^{(1)},\cr
\dot{q}_i(T) + i a_i(q(T)) &= -i p_i^{(2)}.
\end{align}
This is unsatisfactory: if $p_i^{(1,2)}$ are real, the boundary conditions violate reality of $q_i$. Such boundary condition are simply inconsistent in the Euclidean path integral over real functions $q^i(\tau)$. This is of course similar to what we encountered a moment ago, because equations of motion are also complex due to the magnetic field coupling $i a_i(q)\dot{q}^i$.

It is clear how to resolve this issue: in addition to replacing $i a_i(q)\dot{q}^i$ by $s a_i(q)\dot{q}^i$ with $s\in\R$, we have to relax reality of $p_i$ at the boundaries, namely replace $p_i^{(1,2)}$ by $i\pi_i^{(1,2)}$. The modified action reads:
\begin{equation}
S_E=\pi^{(1)}_i q^i(0)- \pi^{(2)}_i q^i(T) +\int\left(\frac12 g_{ij}(q)\dot{q}^i\dot{q}^j +s g_{ij}(q) a^i(q) \dot{q}^j -\frac12 g_{ij}a^i(q)a^j(q) + b(q)\right)\dd\tau.
\end{equation}
Now the EOMs are real, and the boundary conditions are real too:
\begin{align}
\dot{q}_i(0) + s a_i(q(0)) &= \pi_i^{(1)},\cr
\dot{q}_i(T) + s a_i(q(T)) &= \pi_i^{(2)}.
\end{align}

After computing the path integral, we have to analytically continue the answer as follows: 
\begin{equation}
s\to i, \quad \pi_i^{(1,2)} \to -i p_i^{(1,2)}.
\end{equation}

We can think of the analytic continuation of $p_i$ in terms of the original phase space $X=T^*Y$ as first enlarging the space by complexifying the fibers of $T^*Y$:
\begin{equation}
\tilde{X}=T^*_\C Y,
\end{equation}
and then picking a different real cycle in each fiber. In this way, instead of the original $X\subset \tilde{X}$ given by $p_i\in \R$, we pick a different symplectic submanifold, $X_E\subset \tilde{X}$ that has the same base $Y$ but $p_i\in i\R$. 

What we learn from this discussion is that if we compute the boundary state as a functional of boundary conditions using the Euclidean path integral, rewriting it in the Lorentzian signature might involve performing additional analytic continuation of the arguments of this functional. In fact, the above picture suggests a geometric description of this analytic continuation: if the phase space of the theory (in Lorentzian time) is $X=T^*Y$, we should consider an enlarged space $\tilde{X}=T^*_\C Y$, which is still equipped with the two-form $\omega$, and its symplectic subspaces $X$ and $X_E$ related by the Wick rotation in the fibers of $T^*Y$. If we have a wave function described as a half-density $\rho(\psi)$ on some Lagrangian submanifold $L_E\subset X_E$, we extend it to a maximal isotropic (with respect to $\omega$) submanifold $\tilde{L}\subset \tilde{X}$, which will intersect $X$ along the Lagrangian submanifold $L\subset X$. In passing from Euclidean to Lorentzian signature, we have to analytically continue $\rho(\psi)$ from $L_E$ to $L$.

In fact, as is also demonstrated in the Appendix \ref{app:osc} for an example of simple harmonic oscillator, one has to perform this analytic continuation to correctly describe gluing of spacetime manifolds. This is because gluing naturally works when the normal direction to the boundary of spacetime has a signature of time. If we forget to analytically continue the boundary state in the way explained above, the gluing integral might end up being divergent, for example.

The procedure of complexification and choosing a different real symplectic slice might remind to some readers of the work due Gukov and Witten \cite{Gukov:2008ve}. To make a closer connection, we could assume that $Y$, the base of $T^*Y$, is itself a real locus of some complex $Y_\C$. Then we could consider $X_\C=T^*(Y_\C)$, a complex symplectic manifold insider of which we pick two different real symplectic slices $X$ and $X_E$. Description of the phase space as a real locus inside some complex symplectic manifold is precisely the framework of \cite{Gukov:2008ve}. Since in this section we only focused on the case $X=T^*Y$, we did not have to use the full structure of such a complex symplectic manifold: it was only sufficient to complexify the fibers of $T^*Y$. It is conceivable, however, that if we were to study more general symplectic manifolds $X$, proper understanding of Wick rotation might have involved the full structure of $X$ as a real slice inside some complex symplectic manifold $X_\C$. It would be interesting to further explore this connection.

\section{Elementary illustrative examples}

In this section, we are going to describe a few very simple examples of gluing. There is no real novelty in these examples, and the only purpose they serve is to make sure that the reader has the right idea of what was explained in the previous section. Also, as we have already mentioned, the quantum mechanical derivation of gluing from the previous section applies in higher dimensions as well, simply because a QFT on space-time $W\times \R$ can be thought of as a quantum mechanics on $\text{Fields}[W]$, the infinite dimensional space of fields on $W$, with $\R$ being the time direction. In all examples of this section, we are going to work in Euclidean space-time of general dimension $n$, remembering about caveats with analytic continuation of boundary wave functions associated to Euclidean signature, as explained in the previous subsection.

\subsection{Scalar fields}

Consider a real scalar field $\phi$ on an n-dimensional Riemannian manifold $M$. The action is: 
\begin{equation}
S = \int_M \dd^nx \sqrt{g} \left(\frac12 \partial_\mu\phi \partial^\mu\phi + V(\phi)\right).
\end{equation}
Let $W\subset \partial M$ be a boundary component of $M$. (We ignore other boundary components for simplicity.) The simplest choice of boundary conditions is $\phi\big|_W =v$, where $v$ is a fixed function on $W$. Each $v\in C^\infty(W)$ describes a Lagrangian submanifold in the phase space on $W$, and taking all possible $v$, we obtain a foliation of this phase space corresponding to a certain polarization. We refer to it as the Dirichlet polarization. Then the path integral computing the boundary state is simply:
\begin{equation}
\Psi_1[v]=\int_{\phi|_W = v} \pD\phi\, e^{-S},
\end{equation}
where the functional $\Psi_1[v]$ can be also written as $\langle v|\Psi_1 \rangle$, i.e., an overlap between the boundary state $|\Psi_1\rangle$ and the state corresponding to the Lagrangian submanifold determined by $v$.  If we have another manifold $N$, such that $\bar{W}\subset \partial N$, with the same scalar field theory living on it, the quantum dynamics generate a dual state $\langle\Psi_2|$ at the boundary $\bar{W}$. It is represented by a functional $\Psi_2^{\vee}[v]=\langle \Psi_2|v\rangle$. We can glue $M$ to $N$ along $W$ by composing these states, which is represented by the gluing QFT:
\begin{equation}
\int \pD v\, \Psi_2^\vee[v] \Psi_1[v].
\end{equation}

Alternatively, we could pick a different polarization, corresponding to the ``momentum'' representation, and fix $\partial_\perp \phi$ at the boundary, with the normal direction taken outwards. We might refer to this as the Neumann polarization. In this case we have to include a boundary term in order to obtain a correct representation of the boundary state as explained in earlier sections:
\begin{equation}
\Psi_1[p]=\int_{\partial_\perp \phi|_W = p} \pD\phi\, e^{ \int_W \phi\partial_\perp \phi\, -\, S}. 
\end{equation}
The gluing works in a similar way, except that we have to include ``$-i$'' in front of $p$ to account for the analytic continuation since $\Psi[p]$ was computed by the Euclidean path integral:
\begin{equation}
\int \pD p\, \Psi_2^\vee[-ip] \Psi_1[-ip].
\end{equation}

Both these examples demonstrate the gluing procedure in the simplest case possible. They do not illustrate the point about symmetries, because the theory is way too simple. Suppose now that we have $N$ real scalar fields, and the potential $V(\phi)$ is such that the action has $O(N)$ symmetry: 
\begin{equation}
S = \int_M \dd^nx \sqrt{g} \left(\frac12\sum_{i=1}^N \partial_\mu\phi^i \partial^\mu\phi^i + V(\phi)\right).
\end{equation}
If we choose the Dirichlet polarization for all fields or, alternatively, the Neumann polarization, in either case we obtain a polarization preserved by $O(N)$. Therefore, the gluing theories would also have $O(N)$ symmetry.

On the other hand, if we choose the Dirichlet polarization for the first $k$ fields, $\phi^i$, $i=1\dots k$, and the Neumann polarization for the remaining $N-k$ fields, this breaks the $O(N)$ symmetry. Now the boundary state is given by: 
\begin{equation}
\Psi_1[v,p]=\int_{\substack{\phi^i\big|_W=v^i,\, i=1\dots k\\ \partial_\perp\phi^j\big|_W=p^j,\, j=k+1\dots N}} \pD\phi\, e^{\int_W \sum_{j=k+1}^N \phi^j\partial_\perp \phi^j - S}.
\end{equation}
Such polarization is preserved only by the $O(k)\times O(N-k)$ subgroup, and this is the symmetry of the gluing theory:
\begin{equation}
\langle\Psi_2|\Psi_1\rangle = \int \prod_{i=1}^k \pD v^i\, \prod_{j=k+1}^N \pD p^j\, \Psi_2^\vee[v,-ip]\Psi_1[v,-ip].
\end{equation}

\subsubsection{Free fields}
Notice that for the free scalars, one can go even further and explicitly evaluate the boundary states $\Psi_1$ and $\Psi_2$ as functionals of boundary conditions in terms of the bulk Green functions. These states are given by the exponentials of quadratic non-local functionals, roughly of the form $\Psi[v]\sim \exp \left[-\int \dd^{n-1}x\, \dd^{n-1}y\, G(x,y) v(x) v(y) \right]$. Such computations were performed for the first time for free scalars on Riemann surfaces (in the context of string theory) in \cite{Morozov:1988xk, Losev:1988ea, Morozov:1988gj, Morozov:1988bu}.

\subsection{Spinor fields}
The next simplest example to look at is the free Dirac spinor whose action is:
\begin{equation}
\label{DirSpinor}
S=\int_M \dd^nx \sqrt{g}\, \tilde\lambda \slashed{D}\lambda,
\end{equation}
where $\lambda$ and $\tilde\lambda$ are independent because we work on a general Euclidean Spin-manifold. Again we consider a boundary component $W\subset \partial M$. Canonical formalism applied to the first order action \eqref{DirSpinor} implies that the phase space on $W$ can be parametrized by fields $\lambda\big|_W$ and $\tilde\lambda\big|_W$, whose Poisson brackets are:
\begin{equation}
\label{DirPoiss}
\{\lambda_\alpha(x), \tilde\lambda_\beta(y)\}_P = (\gamma_\perp)_{\alpha\beta}\delta_W(x-y),
\end{equation}
where $\delta_W$ is the delta-function on $W$, and $\gamma_\perp=\gamma_\mu e^\mu_\perp$ is the Dirac gamma matrix in the direction of the unit outward normal to $W$. We can choose to fix $\lambda$ at the boundary, in which case the path integral representation of the boundary state is simply:
\begin{equation}
\Psi_1[\rho]=\int_{\lambda\big|_W =\rho} \pD\lambda \pD\tilde\lambda\, e^{-S},
\end{equation}
and the gluing is done via:
\begin{equation}
\int \pD \rho\, \Psi_2^\vee[\rho] \Psi_1[\rho].
\end{equation}
Alternatively, we could fix $\tilde\lambda$ at the boundary. This requires including a boundary term in the path integral representation of the state:
\begin{equation}
\Psi_1[\eta] = \int_{\tilde\lambda\big|_W=\eta } \pD\lambda \pD\tilde\lambda\, e^{\int_W \tilde\lambda \gamma_\perp \lambda -S },
\end{equation}
and the gluing works in a similar way:
\begin{equation}
\int \pD\eta\, \Psi_2^\vee[\eta] \Psi_1[\eta],
\end{equation}
where one should also note that, unlike in the bosonic case, no factors of $i$ are needed.

We could also fix a more general linear combination of $\lambda$ and $\tilde\lambda$ at the boundary, but it would never involve normal derivatives because the equations of motion for $(\lambda, \tilde\lambda)$ are of the first order. Generalization to N spinor fields invariant under $U(N)$ global rotations and the observation how polarization can break it down to various subgroups describing possible symmetries of the gluing theory works in analogy with the scalar field case.

\subsection{A pure gauge theory}\label{sec:puregauge}
Now let us consider a pure Yang-Mills theory (based on a compact gauge group $G$) living on $M$, with the standard action:
\begin{equation}
S={\rm tr}\int_M \frac12 F\wedge *F.
\end{equation}
Applying canonical formalism to this theory is a well-known exercise. Canonical momentum is:
\begin{equation}
\pi^\mu = F^{0\mu}.
\end{equation}
This implies a constraint:
\begin{equation}
\pi^0=0,
\end{equation}
that further implies a secondary constraint:\footnote{Secondary constraints are derived by taking Poisson brackets of the naive Hamiltonian with the constraints that we already have.}
\begin{equation}
D_i \pi^i=0.
\end{equation}
These two are the first class constraints; they determine a constraint surface (that we denoted $C$ before) and generate the gauge symmetry acting on it. It is quite common to complete this by the gauge-fixing conditions:
\begin{equation}
A_0=0,\quad D^i A_i=0.
\end{equation}
The physical Hamiltonian is:
\begin{equation}
H={\rm tr}\left(\frac12 \pi_i \pi^i + \frac14 F_{ij} F^{ij}\right),
\end{equation}
and the phase space action is:
\begin{equation}
S = {\rm tr}\int \dd^n x \left(\pi^i\partial_0 A_i - H\right).
\end{equation}

In the standard Dirichlet polarization, we fix gauge field $A_i$ at the boundary $W\subset \partial M$:
\begin{equation}
A_i\big|_W = a_i.
\end{equation}
The boundary state can either be described in a gauge-fixed or in a gauge-invariant way. In the former case, it is described as a functional of $a_i$ defined on a subspace of fields satisfying $D^i a_i=0$. In the latter case, we drop the gauge-fixing condition $D^i A_i=0$ (like in the general discussion of gauge theories in Section \ref{gauge_general}), and describe the state by a gauge-invariant functional of an unconstrained $a_i$.

Let us make a few remarks about $A_\perp$, the normal component of the gauge field close to the boundary. Its most basic property is that it does not carry any physical data, and corresponds ot a purely gauge degree of freedom in the theory. Therefore, the actual condition that $A_\perp$ satisfies at the boundary depends on the gauge-fixing condition that we pick in the bulk. 

There are many ways to understand this fact. Most fundamentally, all physical information about the boundary condition is encoded in the state it produces. The boundary state, as a wave functional, only depends on $A_i$, not on $A_\perp$, which is also seen from the way we describe the phase space of the gauge theory: we have explained above that the classical description of the phase space involves $A_0=0$, the temporal gauge. In temporal gauge, the gauge field component normal to the spacial slice is zero, which implies that in this description, $A_\perp\big| =0$ is a natural boundary condition. Indeed, it is easy to see that $A_\perp$ can always be gauged away at the boundary.

On the other hand, it was pointed out recently \cite{Witten:2018lgb} that if we use Lorenz gauge in the bulk, then by restricting it to the boundary, one finds that $\cD_\perp A_\perp\big| =-D^i A_i\big|$, or if $A_i\big|=0$ then $\cD_\perp A_\perp\big|=0$. This simply shows once again that the actual condition on $A_\perp$ depends on the gauge. In particular, if needed so for some reason (and especially since we are making contact with the canonical formalism, where the temporal gauge is most natural), we can always assume that we have picked the temporal gauge in the neighborhood of the boundary, and therefore $A_\perp\big| =0$ holds. On the other hand, the gauge choice does not affect the boundary state $\Psi[A_i]$, and hence the gluing procedure we are about to discuss does not depend on the boundary condition on $A_\perp$. Quite often, the bulk calculations are most conveniently formulated in the Lorenz gauge. Therefore, for practical purposes of evaluating $\Psi[A_i]$, it is often most useful to choose the Lorenz gauge, in which case $A_\perp$ would satisfy the Neumann boundary condition.

The boundary state is given by the usual path integral formula (after integrating out $\pi^i$ from the phase space path integral):
\begin{equation}
\Psi_1[a] = \int_{\substack{ A_\parallel|_W=a}} \pD A\, e^{-{\rm tr}\int_M \frac12 F\wedge *F}.
\end{equation}
To glue a state $\Psi_1[a]$ with the dual state $\Psi_2^\vee[a]$ on $\bar{W}\subset \partial N$, we write:
\begin{equation}
\langle\Psi_2|\Psi_1\rangle=\int \pD a\, \Psi_2^\vee[a] \Psi_1[a],
\end{equation}
which by itself is a gauge theory on $W$. This is the ``gluing by gauging at the boundary'' explained in Section \ref{gauge_general}.

What if instead we choose the Neumann polarization? It means that we would like to impose:
\begin{equation}
F_{\perp i}\big|_W = v_i.
\end{equation}
In the abelian case, this condition is gauge-invariant. The boundary wave function is simply a functional of this $v$, and the gluing theory does not have any further gauge symmetries:
\begin{equation}
\label{neum_gauge}
\langle\Psi_2|\Psi_1\rangle=\int \pD v\, \Psi_2^\vee[-i v] \Psi_1[-i v],
\end{equation}
where the origin of $-i$ is the same as before (the analytic continuation of momenta variables).

The non-abelian case is trickier: a boundary condition with generic $v$ breaks gauge symmetry at the boundary to a stabilizer of $v$. The gluing theory can still be formally written as \eqref{neum_gauge}, and consistency implies that the integrand of \eqref{neum_gauge} is gauge-invariant, with the gauge symmetry being the adjoint action:
\begin{equation}
v_i(x) \to g(x) v_i(x) g^{-1}(x),\quad g\in {\rm Hom}(M,G).
\end{equation}
Each $v$ belongs to a gauge orbit isomorphic to ${\rm Hom}(M,G)/{\rm Stab}_v$, and in the end one should perform gauge-fixing and integrate over the space of such orbits in the gluing integral.

Note also that to properly generate the boundary state in this case, we have to include the boundary term, and the wave function is given by the following path integral formula:
\begin{equation}
\Psi_1[v]=\int_{\substack{F_{\perp i}|_W=v_i}} \pD A\, e^{{\rm tr}\int_W A^iF_{\perp i} -{\rm tr}\int_M \frac12 F\wedge *F}.
\end{equation}
Of course, in order to evaluate this path integral, one has to fully fix gauge in the bulk, but we did not make it explicit in the above equation. 

The boundary term $\int_W A^i F_{\perp i}=\int_W A\wedge *v$ might appear to violate gauge-invariance. To understand why it does not, let us focus on the abelian case (which is the same as the non-abelian case for generic $v$, since $v$ then explicitly breaks gauge symmetry at the boundary). Recall that the boundary terms were justified using canonical formalism. For gauge fields, the canonical form of the relevant boundary term is $\int_W A\wedge *\pi$, where $\pi_i$ is the canonical momentum. In the gauge-fixed description of the phase space, as outlined above, we have $\dd_W^* A\big|_W=\dd_W^*\pi\big|_W=0$, where $\dd_W$ is the exterior derivative along $W$. This means that the boundary value $v$ should satisfy the constraint:
\begin{equation}
\dd_W^* v=0.
\end{equation}
By writing $V=*v$, this constraint simply says $\dd_W V=0$. 

The boundary term then becomes $\int_W A\wedge V$. If we relax the gauge condition $\dd_W^* A=0$ and allow $A$ to fluctuate at the boundary, the term $\int_W A\wedge V$ is clearly invariant under small gauge transformations. It is not invariant under large gauge transformations shifting $A$ by elements of $H^1(W,\Z)$. The latter would produce a factor of 
\begin{equation}
\Psi_1[-iv]\propto \sum_{[\omega]\in H^1(W,\Z)} e^{-iQ([\omega],[V])}
\end{equation}
in the wave function, where $Q: H^1 \times H^{d-2}\to \R$ is the intersection pairing, and $d-1=\dim W$. It would be quite interesting to explore this type of gluing more.

\subsubsection{Free fields}
As is always the case, for free fields one can go much further and actually compute the boundary states explicitly. Here we just note that the case of a free Maxwell theory was approached along these lines recently in \cite{Blommaert:2018oue}, where they also considered the non-abelian Yang-Mills theory in two dimensions (and made attempts at generalizing to non-abelian theories in arbitrary dimensions, evaluating the boundary state approximately by truncating, in their language, the bulk-boundary interactions).

\subsection{Gluing by anomalous theory and the inflow}\label{sec:anomglue}
Now we would like to present an example of consistent gluing where the gluing theory appears to have a gauge anomaly. Consider a three-dimensional $U(1)$ gauge theory with $N$ fermions of charge $1$, meaning that we have $N$ fields $\psi_{i\alpha}$ $i=1,\dots, N$ of charge $1$ and $N$ fields $\bar\psi_{i\alpha}$ $i=1,\dots, N$ of charge $-1$. A part of the kinetic term that determines Poisson brackets of the fermions is given by:
\begin{equation}
i\sum_{i=1}^N\varepsilon^{\alpha\beta}\bar\psi_{i\alpha}\gamma^\perp_\beta{}^\sigma \cD_\perp \psi_{i\sigma}.
\end{equation}
The Poisson bracket is $\{\psi_{i\alpha}, \bar\psi_{j\beta}\}\sim \delta_{ij}\gamma^\perp_{\alpha\beta}$, where $\gamma^\perp_\alpha{}^\beta=\sigma_3$, the Pauli matrix, so that $\gamma^\perp_{\alpha\beta}=\sigma_1$. It is consistent to choose $\psi_{i1}$ and $\bar\psi_{i1}$ as Poisson-commuting variables determining the polarization. Therefore, we choose the boundary conditions for the fermions as:
\begin{equation}
\psi_{i1}\big|=\chi_i,\quad \bar\psi_{i1}\big|=\bar\chi_i,
\end{equation}
and we notice that from the two-dimensional point of view, these $\chi_i$, $\bar\chi_i$ are chiral fermions. In addition, we choose the standard Dirichlet polarization for the gauge field, i.e., the boundary condition is:
\begin{equation}
A_\parallel\big| =a.
\end{equation}
Then the boundary states become functionals of $a$, $\chi$ and $\bar\chi$:
\begin{equation}
\Psi[a,\chi,\bar\chi],
\end{equation}
so that the gluing of two states is represented by:
\begin{equation}
\int\pD a\, \pD \chi\, \pD\bar\chi\, \Psi_2^\vee[a,\chi,\bar\chi]\Psi_1[a,\chi,\bar\chi].
\end{equation}
This is the path integral of the two-dimensional gauge theory whose field content consists of the $U(1)$ gauge field $a$ and $N$  charge-1 chiral fermions $\chi_{i}$, $\bar\chi_i$. Such a 2D theory would normally be anomalous due to its unbalanced chiral matter. However, it is well-defined as the gluing theory, with the gauge anomaly canceled by the inflow from the 3D bulk. This cancellation was explicitly shown within the quantum-mechanical approach of the previous section, and we will always assume that once the full ($d$-dimensional) theory is well-defined and non-anomalous, the ($(d-1)$-dimensional) gluing theory is also well-defined, with all apparent anomalies canceled through the inflow from the bulk.

In particular, it means that only the full integration measure, 
\begin{equation}
\pD a\, \pD \chi\, \pD\bar\chi\, \Psi_2^\vee[a,\chi,\bar\chi]\Psi_1[a,\chi,\bar\chi],
\end{equation}
is gauge-invariant, while  $\Psi_1[a,\chi,\bar\chi]$, $\Psi_2^\vee[a,\chi,\bar\chi]$ taken separately, or even their product $\Psi_2^\vee[a,\chi,\bar\chi]\Psi_1[a,\chi,\bar\chi]$, are not gauge invariant \emph{as functionals}. The latter means that they are not annihilated by the Lie derivative implementing gauge transformations on the space of fields. Note that this does not contradict gauge-invariance of states! Both $|\Psi_1\rangle$ and $\langle\Psi_2|$ are gauge-invariant as states, in the sense that they are annihilated by the Gauge law constraint. The Gauss law constraint is represented by a certain differential operator acting on the wave functional, which is \emph{not} the same as the Lie derivative implementing gauge transformations on the space of fields.

\subsection{Gluing despite global anomaly}\label{sec:globanom}
Finally, we give an example of gluing in which the gluing theory appears to have a global anomaly.

Consider a 2D $U(1)$ gauge field coupled to a single Dirac fermion of gauge charge $1$. Such a theory is completely well-defined. Now suppose we put it on a cylinder $S^1\times\R$, and cut it open along $S^1\times \{0\}$. We can then glue it back. By choosing the Dirichlet polarization for the gauge field and a proper gauge-invariant polarization for the fermion, we find that the gluing theory is a 1D $U(1)$ gauge theory on $S^1$ with one Dirac fermion of gauge charge $1$.

Such a 1D theory has a global $\Z_2$ gauge anomaly. More generally, if the fermion has a charge $N\in\Z$, then the global anomaly is present for odd $N$. However, we expect the gluing theory to be completely well-defined, simply because the original 2D theory was consistent. Thus we conclude that the full integration measure of the gluing theory, $\pD \bB\, \langle \Psi_1|\bB\rangle\langle\bB|\Psi_2\rangle$, should be gauge-invariant: the bulk contribution ought to cancel the apparent global anomaly on $S^1$.

\section{Discussion and future directions}
The goal of this paper was to discuss some general properties of the gluing law, which is at the heart of the so-called ``Segal's approach'' to quantum field theory, or ``functorial field theory''. We only gave a few trivial examples of gluing, with the purpose to illustrate general statements made earlier in the paper. The main take-home messages are of somewhat conceptual nature and include: the notion of ``polarized boundary conditions'' for a given polarization on the phase space; understanding that gluing in a Lagrangian theory $\qft_d$ is represented as an integral over a space of polarized boundary conditions (which we treat somewhat formally, only focusing on its local geometry); the idea that this gluing integral can be thought of as a $\qft_{d-1}$ on its own, for which the Main Lemma provides sufficient conditions to acquire a symmetry induced from the symmetry of the parent theory $\qft_d$.

More concrete and less trivial applications of our Main Lemma are discussed in a separate paper \cite{glue2}, where gluing in a supersymmetric $\qft_d$ is represented by a supersymmetric $\qft_{d-1}$ that is solvable by localization.

With bigger goals in mind, we merely scratched the surface of the gluing idea. Moreover, everything said so far only applied to field theories admitting Lagrangian descriptions. While this is an extremely large class of theories a lot of which are well-studied and lead to interesting phenomena in physics and mathematics, there also exists a large number of local field theories of interest that do not have known Lagrangians, and hence cannot be defined using the path integral. Nevertheless, being \emph{local} implies that they still satisfy the gluing property. It just cannot be formulated as an integral over polarized boundary conditions, simply because there is no obvious semiclassical description.

There is a number of generalizations and further directions related to matters discussed here that would be interesting to address in the future. They include:

\begin{itemize}
	\item Generalizations to manifolds with corners. It is very natural to study objects of all possible codimensions in QFT, and one way they can enter the theory is by making arbitrary cuts that produce corners. For example, if we take an ordinary manifold with boundary and make a cut that goes through the boundary, we create a codimension 2 corner. This is an admissible operation for local $\qft_d$, and a similar gluing property should hold for such more general geometries. However, it would be wrong to expect that one can perform cutting and gluing purely within the ``gluing theory'' $\qft_{d-1}$, simply because it is non-local. Rather, it is a property of the local theory $\qft_d$.
	
	To have a sensible cutting and gluing prescription for manifolds with corners, we should always use local boundary conditions. Then we proceed in the following way. First, given a manifold $M$ with boundary $N=\partial M$, we choose some boundary condition at $N$ and treat it as the fixed object -- while before we were thinking of $N$ as supporting some state in $\cH_N$ described by a functional of varying boundary conditions, now we refer to $N$ as supporting a brane, which precisely means that we have fixed the boundary condition. Assuming that this boundary condition is \emph{local} implies that we can perform a further cut of $M$ along a codimension one surface $Y$, possibly resulting in corners. This $Y$ supports some state, and the space of such states $\cH_Y$ might in principle depend on the boundary conditions we fixed in the previous step. After this, we can ``fix a brane'' on $Y$ and perform a further cut etc. The gluing is done in the opposite order: at each step we simply integrate over polarized boundary conditions along the gluing surface, with all other boundaries supporting branes.
	
	In this way, we can, at least in principle, describe theories on triangulated manifolds by gluing them from simplices. If $M$, $\dim(M)=n$, is triangulated, and $M_{(k)}$ denotes is $k$-skeleton, we can break gluing into steps according to codimension: first we integrate over boundary conditions at codimension one cells (i.e. at $M_{(n-1)}\setminus M_{(n-2)}$), with boundary conditions having fixed values along $M_{(n-2)}$, this integration is interpreted as our $\qft_{d-1}$; then we integrate over values of boundary conditions at codimension two cells (i.e. at $M_{(n-2)}\setminus M_{(n-3)}$) with fixed values at the codimension three, this integration is interpreted as $\qft_{d-2}$; then we continue to integrate over values of boundary conditions in codimension three etc.
	
	It would be interesting to exhibit this (or similar) cascade of integrations in some concrete examples, for example in Chern-Simons theory with compact group, where it has a potential to produce a state-integral model. It would also be instructive to follow this procedure for the complex Chern-Simons theory and compare the results with the known state-integral model description studied in a number of papers \cite{2012arXiv1208.1663G,2013arXiv1303.5278G,2014arXiv1409.1208E,2007JGP....57.1895H,Dimofte:2014zga,Dimofte:2009yn,Dimofte:2011gm,Dimofte:2012qj,Andersen:2011bt,2013arXiv1305.4291E,Dimofte:2013iv,Dimofte:2015kkp,Dimofte:2011ju}.
	
	\item Generalization to complex polarizations. In this paper, mostly for simplicity purposes, we chose to focus on real polarizations. As is well-known, one can quantize in complex polarizations, such as holomorphic polarization that is widely applied for K\"ahler phase spaces. The gluing procedure can certainly be described in this case too, with the integral going over some middle-dimensional cycle in the appropriate space of complex boundary conditions. Precise understanding of this case, and in particular how to extent the Main Lemma to complex gluings, would be quite valuable, because it would allow to derive new representations of gluing that are not covered by the formalism of this and companion paper.

	\item Broadly speaking, it would be desirable to understand whether it is possible to obtain any concrete results from gluing in non-topological theories outside the context of symmetries discussed in this and companion paper. While supersymmetry of the gluing theory proves to be very useful for writing concrete non-perturbative finite-dimensional gluing formulas in \cite{glue2}, it just adds another trick to a number of those already available in SUSY theories. Therefore, generalizations to broader classes of theories would be highly valuable.
	
	\item As already mentioned before, the formalism of this paper only applies to a certain class of Lagrangian theories. Strictly speaking, we should only expect it to work, at least in the current form, in renormalizable theories that do not require adding higher-derivative counterterms. It would be interesting to relax any of these restrictions, for example study theories with strongly-interacting UV fixed points (such as five-dimensional gauge theories). Another generalization is to go beyond the two-derivative actions, which would require a more general version of the canonical formalism---the Ostrogradsky formalism. Understanding how much of this can be extended to local \emph{effective} field theories, which is clearly beyond the scope of the current paper, is also a question of some interest.
	
	\item Last but not least, extension to non-Lagrangian theories is in high demand for many topics of current interest. As mentioned earlier in this section, such theories, when they are local, definitely satisfy the gluing law. However, to understand it any better than just a formal inner product on the Hilbert space, one would require a more concrete description of boundary conditions and hence boundary states in non-Lagrangian theories.
\end{itemize}

There is also a number of potential supersymmetric applications and generalization apart from those addressed in the companion paper; they are listed in the Conclusions section of \cite{glue2}.

\acknowledgments

The author thanks Jorgen E. Andersen, Tudor Dimofte, Yale Fan, Bruno Le Floch, Davide Gaiotto, Sergei Gukov, Alexei Morozov, Silviu Pufu, Gustavo J. Turiaci, Ran Yacoby for discussions. The author also thanks Tudor Dimofte and Davide Gaiotto for comments on the draft. This work was supported by the Walter Burke Institute for Theoretical Physics and the U.S. Department of Energy, Office of Science, Office of High Energy Physics, under Award No.\ DE-SC0011632, as well as the Sherman Fairchild Foundation.

\appendix
\section{Path integral derivation of the Main Lemma}\label{appen:main}
In this appendix we would like to rederive the statement about symmetries of the gluing theory purely from the path integral point of view. In the main text we have used a more compact argument, but it is somewhat instructive to see how this works in the phase space path integral.

The path integral measure for the phase space formulation is usually written formally as $\pD y=\prod_t {\rm Pf}\left[ \omega\left(y(t)\right)\right]$. Each factor is the canonical symplectic volume on $X$ and is invariant under symplectomorphisms. We assume that $\pD y$ is also invariant in the bulk of the interval (for $t>0$), i.e., that the symmetry generated by $V_\Phi$ (the Hamiltonian vector field for $\Phi\in C^\infty(X)$) is non-anomalous in the path integral.\footnote{If it could be anomalous, we would restrict only to the class of such $\Phi$ that generate non-anomalous symmetries. However, it is believed that symplectomorphisms are non-anomalous in quantum mechanics, whereas covariance with respect to more general diffeomorphism can be broken by the anomaly \cite{Dedushenko:2010zn}.} 

However, one should be more careful with the definition of this measure, and in particular with the regularization. In the simplest case of symplectic form $\sum_i \dd p_i\wedge \dd q^i$, the standard definition (and the one following from the connection to operator formulation of quantum mechanics) uses lattice regularization: the time interval $[-T,0]$ gets discretized into a set of points $t_n=-T + \frac{n}{N}T$, $n=0..N$. Then one assigns a variable $q^i_{(n)}$ to each site $t_n$ of the lattice and a variable $p^{(n)}_i$ to each interval $(t_{n-1}, t_n)$. The action is written as:
\begin{equation}
S_N = \sum_{n=1}^N \left[p^{(n)}_i (q_{(n)}^i-q_{(n-1)}^i) - H(p_n,q_n)\frac{T}{N} \right].
\end{equation}
This regularization is known to work well, however it makes the action of symplectomorphism $V_\Phi$ somewhat non-trivial. Indeed, a general canonical transformation is a diffeomorphism that mixes position coordinates and momenta, however in such a regularization, positions live on the sites of a lattice while momenta live on the links. Furthermore, if we choose the standard boundary conditions and fix positions at the endpoints:
\begin{equation}
q_{(0)}^i=a^i,\quad q_{(N)}^i=b^i,
\end{equation}
then we have $N$ sets of momenta variables $p^{(n)}_i$, $n=1..N$, and only $N-1$ independent sets of position variables $q_{(n)}^i$, $n=1..N-1$.

We will present one way of dealing with this issue, which works at least for the problem at hands. We will not attempt understanding it in any more generality, such as in relation to other regularizations etc. This is left for brave minds willing to attack the problem of phase space path integral in the future. So, what we do is as follows. For any transformation $\delta p= \epsilon X(p,q), \delta q = \epsilon Y(p,q)$, we write the discrete version as:
\begin{equation}
\delta p^{(n)} = \epsilon \frac{1}{2}\left(X(p^{(n)},q_{(n-1)}) + X(p^{(n)}, q_{(n)})\right),\quad \delta q_{(n)} = \epsilon\frac{1}{2}\left(Y(p^{(n)},q_{(n)})+Y(p^{(n+1)},q_{(n)})\right).
\end{equation}
in other words, each time when we do not know which variable out of two to pick, we take the average of the two possible expressions. For example, since there is no preferred choice whether momentum $p^{(n)}$ should transform and mix with coordinates on the left, i.e. $q_{(n-1)}$, or on the right, i.e. $q_{(n)}$, we simply take the average of the two possible transformations. In particular, for our symplectomorphism generated by
\begin{equation}
\Phi = P_i f^i(Q) + g(Q),
\end{equation}
the transformations are:
\begin{align}
\label{average_transform}
\delta Q^i_{(n)}&=\epsilon f^i(Q_{(n)}),\cr
\delta P_i^{(n)}&=-\epsilon\left[P_j^{(n)}\frac12 (\partial_i f^j(Q_{(n-1)}) + \partial_i f^j(Q_{(n)})) + \frac12 (\partial_ig(Q_{(n-1)}) + \partial_i g(Q_{(n)})) \right].  
\end{align} 
Suppose that we are fixing $Q$ at both ends, that is $Q_{(0)}$ and $Q_{(N)}$ are not integrated over. Then the measure can be written, up to a normalization constant, as follows:
\begin{equation}
\dd^m P^{(1)} \dd^m Q_{(1)} \dd^m P^{(2)} \dd^m Q_{(2)} ... \dd^m Q_{(N-1)} \dd^m P^{(N)}.
\end{equation}
Applying infinitesimal transformation \eqref{average_transform} generates Jacobian factors which we write for $\dd^m Q_{(n)}$ as:
\begin{equation}
\left(1+\epsilon\frac12 \partial_i f^i(Q_{(n)})\right)\dd^m Q_{(n)}\left(1+\epsilon\frac12 \partial_i f^i(Q_{(n)})\right)
\end{equation}
and for $\dd^m P^{(n)}$ as:
\begin{equation}
\left(1-\epsilon\frac12 \partial_i f^i(Q_{(n-1)})\right)\dd^m P^{(n)}\left(1-\epsilon\frac12 \partial_i f^i(Q_{(n)})\right).
\end{equation}
Overall, we get:
\begin{align}
&\left(1-\frac{\epsilon}2 \partial_i f^i(Q_{(0)})\right)\dd^m P^{(1)}\left(1-\frac{\epsilon}2 \partial_i f^i(Q_{(1)})\right) \left(1+\frac{\epsilon}2 \partial_i f^i(Q_{(1)})\right)\dd^m Q_{(1)}\cr &\left(1+\frac{\epsilon}2 \partial_i f^i(Q_{(1)})\right)
\left(1-\frac{\epsilon}2 \partial_i f^i(Q_{(1)})\right) \dd^m P^{(2)}\left(1-\frac{\epsilon}2 \partial_i f^i(Q_{(2)})\right)  ... \left(1+\frac{\epsilon}2 \partial_i f^i(Q_{(N-1)})\right)\cr
&\dd^m Q_{(N-1)} \left(1+\frac{\epsilon}2 \partial_i f^i(Q_{(N-1)})\right) \left(1-\frac{\epsilon}2 \partial_i f^i(Q_{(N-1)})\right) \dd^m P^{(N)} \left(1-\frac{\epsilon}2 \partial_i f^i(Q_{(N)})\right).
\end{align}
All factors inside the product cancel out (up to $O(\epsilon^2)$ which is neglected at the infinitesimal level anyways), and the only ones that remain are:
\begin{equation}
\left(1-\frac{\epsilon}2 \partial_i f^i(Q_{(0)})\right)\left(1-\frac{\epsilon}2 \partial_i f^i(Q_{(N)})\right)=e^{-\frac{\epsilon}2 \partial_i f^i(Q_{(0)})-\frac{\epsilon}2 \partial_i f^i(Q_{(N)})}+ O(\epsilon^2).
\end{equation}
We assume that in the continuum limit, the action transforms in the standard geometric way. Therefore, we expect the following equality to hold:
\begin{equation}
\int_{\begin{matrix}
	Q(-T)=b\cr
	Q(0)=a
	\end{matrix}} \pD P\pD Q\, e^{iS}=e^{-\frac{\epsilon}2 \partial_i f^i(a)-\frac{\epsilon}2 \partial_i f^i(b)}\int_{\begin{matrix}
	Q'(-T)=b'\cr
	Q'(0)=a'
	\end{matrix}} \pD P'\pD Q'\, e^{iS'},
\end{equation}
where $a = a' + \epsilon f(a')$, and the same for $b$. As usual, we can ignore what is going on at the $t=-T$ endpoint of the interval and simply focus on the $t=0$ boundary. Also, we can go back to our previous level of generality and consider an arbitrary Lagrangian submanifold as the boundary condition. Then equality becomes:
\begin{equation}
\int_{y(0)\in \cL(x)}\pD y\, e^{iS}(\dots) = e^{-\frac{\epsilon}{2}\partial_i f^i(x)}\int_{y'(0)\in [\cL(x)]'}\pD y'\, e^{iS'}(\dots),
\end{equation}
where $[\cL(x)]'$ is the image of $\cL(x)$ under $x = x' + \epsilon f(x')$, and $S'$ is the transformed action. 
\subsection{Gluing and symmetries}

Looking at how the action transforms, we see that the $-H \dd t$ term is invariant, while the transformation of the boundary term $F|_{t=0}$ is simply $\delta F = \epsilon X_\Phi (F)$, and:
\begin{equation}
\delta\alpha = \epsilon \cL_{X_\Phi}\alpha = \epsilon(\dd \alpha(X_\Phi) + \iota_{X_\Phi}\omega) = \epsilon \dd\left(\alpha(X_\Phi) - \Phi \right).
\end{equation}
So we can write:
\begin{equation}
\int_{y'(0) \in [\cL(x)]'}\pD y'\, e^{iS'}(\dots) = \int_{y(0)\in \cL(x')}\pD y\, e^{iS + i\epsilon (X_\Phi^\mu \partial_\mu F + \alpha(X_\Phi) - \Phi)|_{t=0}}(\dots).
\end{equation}
If we recall that locally $\dd F + \alpha = P_i \dd Q^i$, the correction term in the action becomes:
\begin{equation}
X_\Phi^\mu \partial_\mu F + \alpha(X_\Phi) - \Phi= (P_i \dd Q^i)(X_\Phi)-\Phi=P_i f^i(Q) - \Phi = -g(Q).
\end{equation}
Assuming that additional insertions $(\dots)$ are invariant under $X_\Phi$, we obtain that:
\begin{align}
\psi_+(x) &= \int_{y(0)\in \cL(x)}\pD y\, e^{iS}(\dots) = e^{-i\epsilon g(x')-\frac{\epsilon}{2} \partial_i f^i(x')}\int_{y(0)\in \cL(x')}\pD y\, e^{iS}(\dots)\cr\cr
&= e^{-i\epsilon g(x')-\frac{\epsilon}{2} \partial_i f^i(x')} \psi_+(x'),
\end{align}
where now $x'= x - \epsilon f(x)$. Taking the term proportional to $\epsilon$, we find:
\begin{equation}
\label{psiplusinv}
-i f^i(x)\partial_i \psi_+(x) + g(x)\psi_+(x) - \frac{i}{2}\partial_i f^i(x)\psi_+(x)=0.
\end{equation}
The latter is just the statement that $\psi_+(x)$ is annihilated by:
\begin{equation}
\widehat{\Phi}=f^i(x)\widehat{p}_i + g(x) - \frac{i}2 \partial_i f^i=\frac12 (f^i \widehat{p}_i + \widehat{p}_i f^i) + g(x).
\end{equation}
This is the natural Hermitian quantum operator associated to $\Phi$, and \eqref{psiplusinv} is the usual property of the theory whose vacuum preserves the global symmetry: path integral with invariant insertions produces an invariant state. If we keep this statement in the form:
\begin{equation}
\psi_+(x) = e^{-i\epsilon g(x')-\frac{\epsilon}{2} \partial_i f^i(x')} \psi_+(x'),
\end{equation}
and note that the same computation for $\psi_-$ (which lives at the endpoint of the opposite orientation) gives:
\begin{equation}
\psi_-(x) =e^{i\epsilon g(x')-\frac{\epsilon}{2} \partial_i f^i(x')} \psi_-(x'),
\end{equation}
we can conclude that:
\begin{equation}
\psi_+(x)\psi_-(x) = e^{-\epsilon\partial_i f^i(x')}\psi_+(x')\psi_-(x').
\end{equation}
On the other hand, the measure on $X/D$, or more concretely on $\cL^\perp$, which is locally given by $\dd^{m}Q$ and is formally written as $\dd^{m}x$, transforms in the opposite way:
\begin{equation}
\dd^{m}x = e^{\epsilon \partial_i f^i(x')}\dd^m x',
\end{equation}
canceling the similar factor in $\psi_+\psi_-$. So $\psi_+(x) \psi_-(x) \dd^m x$ gives an invariant measure on $X/D$ with the symmetry $x \to x - \epsilon f(x)$ descending from $V_\Phi$ acting on $X$ by the polarization-preserving symplectomorphisms: 
\begin{equation}
\psi_+(x)\psi_-(x)\dd^m x = \psi_+(x')\psi_-(x')\dd^m x'.
\end{equation}
In other words, the $\qft_0$ that describes gluing and is given by the integral:
\begin{equation}
\int_{X/D} \psi_-(x) \psi_+(x),
\end{equation}
has an induced symmetry if the parent theory $\qft_1$ has a symmetry $V_\Phi$ that preserves polarization used in the definition of cutting/gluing. 

Moreover, for this property to hold, there is one more condition: quantum states $\psi_\pm \in \cH$ should be annihilated by the conserved charge $\widehat{\Phi}$. In the above, this was not made fully explicit and was almost automatic: we simply assumed that whatever happens at $t=-T$, preserves the symmetry, and then $\psi_+$ also does, as manifested by the equation \eqref{psiplusinv}. If there is a state $\psi_{in}\in\cH$ which we feed into the path integral as a boundary condition at $t=-T$, that state has to satisfy $\widehat{\Phi}|\psi_{in}\rangle=0$. If we take $T$ to infinity instead, then we should assume that at the infinite past, the system was in the vacuum state $|0\rangle$, and the symmetry is preserved by the vacuum: $\widehat{\Phi}|0\rangle=0$.

While the latter is certainly true for continuous symmetries in our 1-dimensional example, it might fail in higher dimensions if the symmetry is spontaneously broken. In this case, the path integral for $\qft_d$ would have a symmetry at the level of fields, but the vacuum would break it at the level of states. In such situations, the gluing theory $\qft_{d-1}$ would not posses an induced symmetry.

\section{Wick rotation in phase space: harmonic oscillator}\label{app:osc}
To illustrate how Wick rotation of boundary conditions discussed in Section \ref{sec:analytic} works, let us consider an explicit example of a simple harmonic oscillator. Using it as a playground, we are going to observe that indeed, wave functions written in the momentum representation undergo additional analytic continuation in passing between Euclidean and Minkowsi signature, which is not present for the position space wave functions. This point, while being quite trivial in the present example, helps to avoid confusions in more complicated cases, such as higher-dimensional theories and more general polarizations. The main lesson it teaches is that gluing works naturally along space-like slices in a space-time of Lorentzian signature, while gluing in Euclidean space-time should be properly understood through the analytic continuation.

With this motivation, consider the action:
\begin{equation}
S_L=\int_0^T \dd t \left( \frac{\dot q^2}{2} - \frac{\omega^2q^2}{2} \right).
\end{equation}
The Hamiltonian formalism action is:
\begin{equation}
S_H=\int_0^T \dd t\left(p \dot{q} - H \right),\quad H=\frac{p^2}{2} + \frac{\omega^2 q^2}{2}.
\end{equation}
The standard (Lorentzian time and position space) transition amplitude is given by:
\begin{align}
\langle x_2| e^{-i TH}|x_1\rangle=\int_{\substack{q(0)=x_1\\ q(T)=x_2}} \pD q\, e^{i S_L} = \int_{\substack{q(0)=x_1\\ q(T)=x_2}} \pD q\,\pD p\, e^{i S_H},
\end{align}
where we give both the configuration space and the phase space path integral expressions. Since the action is quadratic, both can be straightforwardly evaluated, with the answer:
\begin{equation}
\label{xxLor}
\langle x_2| e^{-i TH}|x_1\rangle=\sqrt{\frac{\omega}{2\pi i \sin\omega T}}\times\exp\left\{\frac{i\omega}{2\sin\omega T}[(x_1^2 + x_2^2)\cos\omega T -2x_1 x_2] \right\}.
\end{equation}
To obtain the momentum representation of the same transition amplitude, we can either perform the Fourier transform:
\begin{align}
\label{ppLor}
\langle p_2| e^{-i TH}|p_1\rangle&=\frac1{2\pi} \int \dd x_1 \dd x_2\, e^{-i p_2 x_2}\langle x_2| e^{-i TH}|x_1\rangle e^{i p_1 x_1}\cr
&=\sqrt{\frac{1}{2\pi i\omega \sin\omega T}}\times\exp\left\{\frac{i}{2\omega\sin\omega T}[(p_1^2 + p_2^2)\cos\omega T -2p_1 p_2] \right\},
\end{align}
or evaluate the path integral with the appropriate boundary conditions and boundary terms:
\begin{align}
\langle p_2| e^{-i TH}|p_1\rangle&=\int_{\substack{p(0)=p_1\\ p(T)=p_2}} \pD q\,\pD p\, e^{ip_1q(0) -ip_2q(T)+i S_H}=(\text{Integrate out $p$})\cr
&=\int_{\substack{\dot{q}(0)=p_1\\ \dot{q}(T)=p_2}} \pD q\, e^{ip_1q(0) -ip_2q(T)+i S_L},
\end{align}
which of course gives the same answer.

Now let us see how the Euclidean time version works. The Lagrangian and Hamiltonian actions in Euclidean time are:
\begin{align}
S_L^E&=\int_0^T\dd\tau\left(\frac{\dot{q}^2}{2} + \frac{\omega^2 q^2}{2}\right),\cr
S_H^E&=\int_0^T\dd\tau \left(-ip\dot{q} + H\right).
\end{align}
The position representation transition amplitude is obtained by the standard Wick rotation:
\begin{align}
\langle x_2| e^{-TH}|x_1\rangle=\int_{\substack{q(0)=x_1\\ q(T)=x_2}} \pD q\, e^{- S_L^E} = \int_{\substack{q(0)=x_1\\ q(T)=x_2}} \pD q\,\pD p\, e^{-S_H^E},
\end{align}
with the answer 
\begin{equation}
\langle x_2| e^{-TH}|x_1\rangle=\sqrt{\frac{\omega}{2\pi \sinh\omega T}}\times\exp\left\{-\frac{\omega}{2\sinh\omega T}[(x_1^2 + x_2^2)\cosh\omega T -2x_1 x_2] \right\},
\end{equation}
that is of course related to \eqref{xxLor} by $T\to iT$. Now let us look at the momentum representation where we fix $p(0)=p_1$ and $p(T)=p_2$. The Hamiltonian formalism action with the appropriate boundary terms, upon Wick rotation, then becomes:
\begin{equation}
S_H^E=i p_2 q(T) - ip_1 q(0) + \int_0^T\dd\tau \left(-ip\dot{q} + H\right).
\end{equation}
This path integral is supposed to give an expression for $\langle p_2| e^{-TH}|p_1\rangle$ that is related to \eqref{ppLor} by a simple Wick rotation $T \to iT$, so that the following is literally true from the operator formalism perspective too:
\begin{equation}
\label{ppEucl}
\langle p_2| e^{-TH}|p_1\rangle=\sqrt{\frac{1}{2\pi \omega \sinh\omega T}}\times\exp\left\{-\frac{1}{2\omega\sinh\omega T}[(p_1^2 + p_2^2)\cosh\omega T -2p_1 p_2] \right\}.
\end{equation}
Now let us try to proceed in steps and first integrate out $p$ to write a Euclidean Lagrangian action with boundary terms. This gives:
\begin{align}
\label{ppLorPI}
\langle p_2| e^{-TH}|p_1\rangle=\int_{\substack{\dot{q}(0)=-ip_1\\ \dot{q}(T)=-ip_2}}\pD q e^{ip_1 q(0) - ip_2 q(T) - \int_0^T \dd\tau \left(\frac{\dot{q}^2}{2} + \frac{\omega^2 q^2}{2} \right)},
\end{align}
which is formally supposed to be correct, however it has one apparent problem: boundary conditions on $\dot{q}$ now break reality of $q$! Certainly, if we were to start from the Euclidean time path integral, we would impose boundary conditions that respect reality of the field variable. While it is acceptable to have classical solutions that violate reality conditions---this simply means that the saddle point lies off the integration cycle, and we are allowed to deform the cycle to make it pass through the saddle point in the standard steepest descent approach---it is certainly safer to avoid reality-breaking boundary conditions. Therefore, we replace $p_a \to i\pi_a$ in the above path integral, to define the following quantity:
\begin{align}
K_E(\pi_1, \pi_2; T)=\langle \pi_2| e^{-TH}|\pi_1\rangle=\int_{\substack{\dot{q}(0)=\pi_1\\ \dot{q}(T)=\pi_2}}\pD q e^{-\pi_1 q(0) + \pi_2 q(T) - \int_0^T \dd\tau \left(\frac{\dot{q}^2}{2} + \frac{\omega^2 q^2}{2} \right)}.
\end{align}
This quantity has a well-defined path integral expression in the Euclidean time, however it is not immediately equal to the Wick-rotated momentum-space amplitude \eqref{ppLor}, \eqref{ppLorPI}. The above discussion shows that the relation is through the additional analytic continuation of the momentum space variable:
\begin{equation}
\langle p_2| e^{-TH}|p_1\rangle = K_E(-ip_1, -ip_2; T).
\end{equation}
Indeed, an explicit evaluation of the path integral shows that:
\begin{equation}
K_E(\pi_1, \pi_2, T)=\sqrt{\frac{1}{2\pi \omega \sinh\omega T}}\times\exp\left\{\frac{1}{2\omega\sinh\omega T}[(\pi_1^2 + \pi_2^2)\cosh\omega T -2\pi_1 \pi_2] \right\},
\end{equation}
and wee see that this expression is related to \eqref{ppEucl} by $\pi_a \to -ip_a$. This is the analytic continuation of the momentum variable discussed in the main text.

After belaboring the point on analytic continuation this much that it becomes completely trivial, let us make a few more comments on physical states. We mentioned in the main text that one can use the Euclidean path integral on a compact manifold $M$ with boundary $N=\partial M$ to generate a state in $\cH_N$. Such a state can be further fed into the path integral in Lorentzian signature as its boundary condition. For example, we can use a state $\psi_M \in \cH_N$ generate by the Euclidean dynamics on $M$ as a boundary condition at $t=0$ on $N\times \R_+$, with a time-like direction $\R_+=(0,\infty)$, to propagate it along $\R_+$. 

\begin{figure}[t!]
	\centering
	\includegraphics[width=0.8\textwidth]{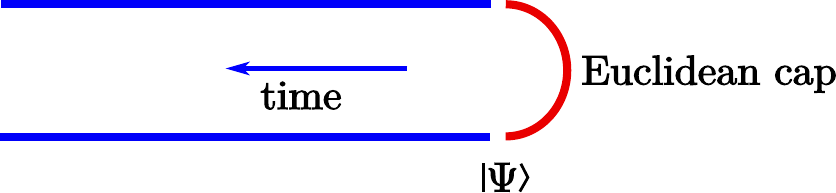}
	\caption{\label{fig:cap} The red region represents a Euclidean ``cap'' that creates a state $|\Psi\rangle$. This state is then propagated in $\qft_1$ on $\R_+\amalg\R_+$ using the Lorentzian evolution $e^{-it H}$. For the oscillator, $H = \frac{p_1^2}{2} + \frac{\omega^2 q_1^2}{2}+\frac{p_2^2}{2} + \frac{\omega^2 q_2^2}{2}$ is acting on $\cH_{\rm pt}\otimes \cH_{\rm pt}$, where $\cH_{\rm pt}$ is a Hilbert space of a single oscillator.}
\end{figure}

Let us observe an example of this in our oscillator toy model. Unfortunately, it is not possible with a single oscillator, because there are no compact one-dimensional manifolds with a single boundary component (to which one would attach the ``time'' $\R_+$), there is always an even number of boundaries. What we can do, however, is consider two copies of an oscillator as each evolving on its own ``time'' manifold $\R_+$. In other words, we take the ``space'' to consist of two disjoint points rather than one, so we consider a 1D QFT on $\R_+ \amalg \R_+$ with Lorentzian metric. Then we connect these two lines by a finite interval with Euclidean metric on it, which we interpret as creating a boundary state for the $\qft_1$ on $\R_+ \amalg \R_+$, see illustration on Figure \ref{fig:cap}.

We can interpret the red cap as a region creating the ``in'' state which we further evolve using Lorentzian dynamics. This state belongs to the space $\cH_{{\rm pt } \amalg {\rm pt}}=\cH_{\rm pt}\otimes \cH_{\rm pt}$. We could either equip the red cap with Lorentzian or Euclidean metric. In the former case, this would create a continuum spectrum-normalizable state (which does not belong to $L^2(\R^2)$). With Euclidean metric, we expect to obtain a square-normalizable state. This is obviously true if we use the position representation to write down this state: 
\begin{equation}
\label{cap_state_x}
\Psi(x_1, x_2)=\sqrt{\frac{\omega}{2\pi \sinh\omega T}}\times\exp\left\{-\frac{\omega}{2\sinh\omega T}[(x_1^2 + x_2^2)\cosh\omega T -2x_1 x_2] \right\},
\end{equation}
where $T$ is the length of the red cap region.

However, in the momentum representation, with the naive boundary condition $\partial_\perp q\big| = (\pi_1, \pi_2)$ (where $\partial_\perp$ is a time derivative pointing outward, i.e. in time direction), one finds:
\begin{align}
\Psi_E(\pi_1, \pi_2)&=\int_{\partial_\perp q|=(\pi_1, \pi_2)}\pD q e^{-S_E}\cr
&=\sqrt{\frac{1}{2\pi \omega \sinh\omega T}}\times\exp\left\{\frac{1}{2\omega\sinh\omega T}[(\pi_1^2 + \pi_2^2)\cosh\omega T +2\pi_1 \pi_2] \right\},
\end{align}
We see that this is not normalizable in any way. Of course, as it should be completely obvious by now, the resolution to preform analytic continuation in momentum:
\begin{align}
\Psi(p_1,p_2)&=\Psi_E(-ip_1, -ip_2)\cr
&=\sqrt{\frac{1}{2\pi \omega \sinh\omega T}}\times\exp\left\{-\frac{1}{2\omega\sinh\omega T}[(p_1^2 + p_2^2)\cosh\omega T +2p_1 p_2] \right\},
\end{align}
where $\Psi(p_1,p_2)$ is perfectly square-normalizable now. 

As a consistency check, this $\Psi(p_1, p_2)$ is nothing else but a Fourier transform of $\Psi(x_1, x_2)$ from \eqref{cap_state_x}, $\Psi(p_1, p_2)=\frac1{2\pi}\int \dd x_1 \dd x_2 e^{-i p_1 x_1 - ip_2 x_2}\Psi(x_1, x_2)$.

Another operation that could be performed on the Euclidean cap from Figure \ref{fig:cap} is gluing to itself, or to another copy of the Euclidean cap. The result, quite expectedly, is the thermal partition function of an oscillator. In the position representation, gluing to itself is simply given by:
\begin{equation}
\int\dd x\, \Psi(x,x)=\frac1{2\sinh \frac{\omega T}{2}},
\end{equation}
which is indeed the thermal partition on $S^1$ of length $T$, while gluing to another copy of the cap is:
\begin{equation}
\int\dd x_1\dd x_2\, \Psi(x_1,x_2)\Psi(x_1, x_2)=\frac1{\sinh\omega T},
\end{equation}
the same partition function but on the circle of length $2T$.

Now if we would like to perform the same gluing in momentum space, we clearly should not use the Euclidean wave function $\Psi_E(\pi_1,\pi_2)$ as it is divergent, instead we use the analytically continued (or Lorentzian) wave function $\Psi(p_1,p_2)=\Psi_E(-ip_1,-i p_2)$. For gluing to itself, we also write $\Psi(p,-p)$ since orientations of the two ends must be opposite to allow gluing:
\begin{align}
\int\dd p\, \Psi_E(-ip, ip)=\frac1{2\sinh \frac{\omega T}{2}},
\end{align}
while for gluing of the two copies we integrate as follows:
\begin{equation}
\int\dd p_1\dd p_2\, \Psi_E(-ip, -ip)\Psi_E(ip, ip)=\frac1{\sinh{\omega T}}.
\end{equation}
The latter example is relevant for what we are doing in the companion paper \cite{glue2}, since we mostly glue Euclidean theories there, and as we emphasize here, to obtain correct answer we really should analytically continue Euclidean wave functions before gluing. This observations explains some otherwise mysterious factors of $i$ in \cite{glue2}.

\newpage

\bibliographystyle{JHEP}
\bibliography{gluing}

\end{document}